\documentclass{elsart}

\usepackage{natbib}

\usepackage{amssymb}

\begin{document}

\begin{frontmatter}

\title{On the verge of {\textit {\textbf Umdeutung} in Minnesota: Van Vleck and the correspondence principle. Part One.}\thanksref{mpiwg}}
\thanks[mpiwg]{This paper was written as part of a joint project in the history of quantum physics of the {\it Max Planck Institut f\"{u}r Wissenschaftsgeschichte} and the {\it Fritz-Haber-Institut} in Berlin.}

\author[duncan]{Anthony Duncan},
\author[janssen]{Michel Janssen\corauthref{cor}}
\corauth[cor]{Corresponding author. Address: Tate Laboratory of Physics, 116 Church St.\ NE, Minneapolis, MN 55455, USA, Email: janss011@tc.umn.edu}
\address[duncan]{Department of Physics and Astronomy, University of Pittsburgh}
\address[janssen]{Program in History of Science, Technology, and Medicine, University of Minnesota}

\begin{abstract}
In October 1924,  {\it The Physical Review}, a relatively minor journal at the time, published a remarkable two-part paper  by John H.\ Van Vleck, working in virtual isolation at the University of Minnesota. Van Vleck combined advanced techniques of classical mechanics with Bohr's correspondence principle and Einstein's quantum theory of radiation to find quantum analogues of classical expressions for the emission, absorption, and dispersion of radiation. For modern readers Van Vleck's paper is much easier to follow than the famous paper by Kramers and Heisenberg on dispersion theory, which covers similar terrain and is widely credited to have led directly to Heisenberg's {\it Umdeutung} paper. This makes Van Vleck's paper extremely valuable for the reconstruction of the genesis of matrix mechanics. It also makes it tempting to ask why Van Vleck did not take the next step and develop matrix mechanics himself. 
\end{abstract}


\end{frontmatter}

\section{Introduction}

It is widely acknowledged among historians of modern physics that the famous {\it Umdeutung} [reinterpretation] paper with which Wer\-ner Hei\-senberg (1901--1976) laid the basis for matrix mechanics \citep{Heisenberg 1925c} grew out of a paper he and Hendrik A.\ (Hans) Kramers (1894--1952) co-authored on dispersion theory \citep{Kramers and Heisenberg 1925}.  Although hardly impartial as one of Kramers' students and his biographer, Max \citet{Dresden} calls the Kramers-Heisenberg paper  ``the direct, immediate, and exclusive precursor to the Heisenberg paper on matrix mechanics" (p.\ 275). Martin J.\ \citet{Klein 1970} is more restrained but does agree that ``this work was the immediate predecessor of Heisenberg's new quantum mechanics" (p.\ 31). To understand the origin of matrix mechanics, one thus has to come to grips with the contents of the Kramers-Heisenberg paper.  According to Jagdish Mehra and Helmut Rechenberg, this paper was written ``in such a way that every physicist, theoretician or experimentalist, interested in the subject could understand" \cite[Vol.\ 2, pp.\ 181]{Mehra Rechenberg}.\footnote{This multi-volume history of quantum physics brings together a wealth of information and we shall frequently refer to it. However, it needs to be used with some caution (see, e.g., notes \ref{plagiarism1}, \ref{plagiarism2}, and \ref{plagiarism3} below as well as the review of the first few volumes by John L.\ Heilbron (1985)).} An uniniated modern reader turning to the Kramers-Heisenberg paper after these encouraging words is likely to be disappointed. The authors assume their readers to be thoroughly familiar with techniques, borrowed from celestial mechanics, for dealing with multiply-periodic systems, including canonical transformations, action-angle variables, and related perturbation methods. As far as their contemporaries in theoretical physics were concerned, this was undoubtedly a reasonable assumption. So, Mehra and Rechenberg are probably right to the extent that the intended audience would have had no special difficulties with the paper. The same cannot be said for most modern readers, who no longer have the relevant techniques at their fingertips. Fortunately, there is another paper from the same period covering some of the same terrain that is much easier to follow for such readers. 

Immediately preceding the translation of \citep{Kramers and Heisenberg 1925} in the well-known anthology on the development of matrix mechanics edited by Bartel Leendert van der Waerden (1903--1996) (1968), there is a paper by the American theoretical physicist John Hasbrouck Van Vleck (1899--1980) (1924b). Like the Kramers-Heisenberg paper, it combines some sophisticated classical mechanics with the correspondence principle of Niels Bohr (1885--1962) and the quantum radiation theory of Albert Einstein (1879--1955).
The last section of this paper provides the first published proof that the Kramers dispersion formula, which \citet{Kramers 1924a, Kramers 1924b} had only presented in two short notes in {\it Nature} at that point, merges with the classical formula in the limit of high quantum numbers. Van Vleck's paper is a paragon of clarity. In an interview by Thomas S. Kuhn for the {\it Archive for History of Quantum Physics} (AHQP) in 1963,\footnote{Between February 1962 and May 1964, about 95 people were interviewed for the project \citep[p.\ 3]{Kuhn et al. 1967}. With one exception (see sec.\ 2.4) the exact dates of these interviews are unimportant for our purposes and will not be given when we quote from the transcripts.} Van Vleck acknowledged the influence of his father, the mathematician Edward Burr Van Vleck (1863--1943), in developing his exceptionally lucid writing style: 
\begin{quotation}
My father got after me for my very poor style of scientific exposition. I feel I owe a great deal to him for his splitting up my sentences into shorter sentences, avoiding dangling participles---i.e., tightening up my prose style---the same kind of drill I try to give my own graduate students now.\footnote{\label{interview}P.\ 21 of the transcript of the first of two sessions of the interview, quoted in \citep[p.\ 57]{Fellows}. Van Vleck is talking specifically about the summer of 1925, when he was working on his book-length \citep{Van Vleck 1926a}, but his father had probably given him a few pointers before. \citep{Van Vleck 1924b} definitely belies the author's harsh judgment of his earlier writing style.}
\end{quotation} 
Van der Waerden only included the quantum part, \citep{Van Vleck 1924b}, of a two-part paper in his anthology. In the second part, \citet{Van Vleck 1924c} clearly laid out the results from classical mechanics needed to understand the first part as well as those parts of \citep{Kramers and Heisenberg 1925}  that are most important for understanding Heisenberg's {\it Umdeutung} paper. This is true even though Van Vleck only covered {\it coherent} scattering---i.e., the case in which the frequency of the scattered radiation is the same as that of the incoming radiation---whereas a large part of the Kramers-Heisenberg paper is devoted to {\it incoherent} scattering, first predicted in \citep{Smekal 1923} and verified experimentally a few years later \citep{Raman 1928, Landsberg and Mandelstam 1928}. In his interview with Kuhn, Heisenberg emphasized the importance of this part of his paper with Kramers for the {\it Umdeutung} paper.\footnote{P.\ 18 of the transcript of session 4 of a total of 12 sessions of the AHQP interview with Heisenberg.\label{raman}} Of course, this is also the one part to which Heisenberg materially contributed.\footnote{According to \citet[pp.\ 273--274]{Dresden}, Kramers added Heisenberg's name to \citep{Kramers and Heisenberg 1925} mainly as a courtesy. For Heisenberg's side of the story, see pp.\ 15--18 of the transcript of session 4 of the AHQP interview with Heisenberg, several passages of which can be found in \citep[Vol.\ 2, pp.\ 178--179]{Mehra Rechenberg}, although  the authors cite their own conversations with Heisenberg as their source (cf. the foreword to Vol.\ 2).\label{plagiarism1}}  Still,  the non-commutative multiplication rule introduced in the {\it Umdeutung} paper may well have been inspired, as Heisenberg suggests, by manipulations in this part of the Kramers-Heisenberg paper. To understand where the arrays of numbers subject to this rule come from, however, it suffices to understand how coherent scattering is treated in Kramers' dispersion theory: indeed, the only explicit use of dispersion theory in the {\it Umdeutung} paper is the result for coherent scattering.

\subsection{On the verge of {\rm Umdeutung}}

As in the case of \citep{Kramers and Heisenberg 1925},  one is struck in hindsight by how close \citep{Van Vleck 1924b, Van Vleck 1924c} comes to anticipating matrix mechanics. 
During the AHQP interview, Kuhn reminded Van Vleck of a remark he had made two years earlier to the effect that, if he had been ``a little more perceptive," he ``might have taken off from that paper to do what Heisenberg did." ``That's true," Van Vleck conceded, but added with characteristic modesty: ``Perhaps I should say {\it considerably} more perceptive."\footnote{See p.\  24 of the transcript of the first session of the interview. Kuhn's recollection is that Van Vleck's earlier remark was made during a meeting  in Philadelphia in February 1961 to plan for the AHQP project \citep[pp.\ vii--viii]{Kuhn et al. 1967}. It was only natural for Van Vleck to get involved in Kuhn's project. As a young physicist right after World War II, Kuhn had worked with Van Vleck  \citep[p.\ 518]{Anderson}, a collaboration that resulted in a joint paper \citep{Kuhn and Van Vleck 1950}.} 
In the biographical information he supplied for the AHQP, Van Vleck noted:
\begin{quotation}
In the two or three years after my doctorate \ldots my most significant paper was one on the correspondence principle for absorption \ldots It was somewhat related to considerations based on the correspondence principle that led Heisenberg to the discovery of quantum mechanics, but I did not have sufficient insight for this.\footnote{Biographical information prepared for the American Institute of Physics project on the history of recent physics in the United States (included in the folder on Van Vleck in the AHQP), p. 1.}
\end{quotation}
This modest assessment  is reflected in the discussion of the relation between Van Vleck's work and matrix mechanics by Fred \citet[pp.\ 74--81]{Fellows}, who wrote a superb dissertation covering the first half of Van Vleck's life and career. In a biographical memoir about his teacher and fellow Nobel laureate, Phil \citet{Anderson}\footnote{Van Vleck, Anderson, and Sir Nevill Mott  shared the 1977 Nobel prize ``for their fundamental theoretical investigations of the electronic structure of magnetic and disordered systems." Van Vleck won for work begun in the early 1930s that earned him the title of ``father of modern magnetism."} is less reserved: ``This paper comes tantalizingly close to the kind of considerations that led to Heisenberg's matrix mechanics" (p.\ 506).

Van Vleck did not pursue his own research any further in 1924 and instead spent months writing---and, as he jokingly put it, being a ``galley slave" \citep[p.\ 100]{Fellows} of---a {\it Bulletin} for the {\it National Research Council} (NRC)  on the old quantum theory \citep{Van Vleck 1926a}.
With his masterful survey he would surely have rendered a great service to the American physics community had it not been for the quantum revolution of 1925--1926. Like the better-known  {\it Handbuch} article by Wolfgang Pauli (1900--1958) (1926), the  {\it Bulletin} was, as \citet{Van Vleck 1971} recognized, ``in a sense \ldots obsolete by the time it was off the press" (p.\ 6).\footnote{For the reception of Van Vleck's {\it Bulletin}, see \citep[pp.\ 88--89]{Fellows}. Van Vleck's {\it Bulletin} and Pauli's {\it Handbuch} article were not the only treatises on the old quantum theory that were out of date before the ink was dry. \citep{Born 1925} and \citep{Birtwistle 1926}, two books on atomic mechanics, suffered the same fate.}
One is left wondering what would have happened, had the young assistant professor at the University of Minnesota continued to ponder the interaction between radiation and matter and the correspondence principle instead of fulfilling his duties as a newly minted member of the American physics community. 

That Kramers and Van Vleck---and, one may add, Max Born (1882--1970)  and Pascual Jordan (1902--1980)---came so close to beating Heisenberg to the punch makes the birth of matrix mechanics reminiscent of the birth of special relativity.  The comparison seems apt, even though none of these authors anticipated as much of the new theory as H.\ A.\ Lorentz (1853--1928) and Henri Poincar\'{e} (1854--1912) did in the case of relativity.\footnote{In his autobiography, \citet[pp.\ 216--217]{Born 1978} exaggerated how close he came to matrix mechanics before Heisenberg.} \citet[p.\ 63]{Heisenberg 1971} himself actually compared his {\it Umdeutung} paper to Einstein's relativity paper \citep{Einstein 1905}.  He argued that what his work had in common with Einstein's was its insistence on allowing only observable quantities into physical theory. The analogy is considerably richer than that. 

The breakthroughs of both Einstein and Heisenberg consisted, to a large extent, in reinterpreting elements already present in the work of their predecessors, extending the domain of application of these elements, and discarding unnecessary scaffolding. Einstein recognized the importance of Lorentz invariance beyond electromagnetism, reinterpreted it as reflecting a new space-time structure, and discarded the ether \citep{Janssen 2002}. In the case of \citep{Heisenberg 1925c}, the element of {\it Umdeutung} or reinterpretation is emphasized in the title of the paper. Heisenberg reinterpreted elements of the Fourier expansion of the position of an electron entering into the construction of the Kramers dispersion formula, discarded the orbits supposedly given by that position, and recognized that the non-commuting arrays of numbers associated with transitions between different states and
representing position in his new scheme were meaningful far beyond the dispersion theory from which they originated. 

A further point of analogy is that neither Einstein nor Heisenberg presented the new theory in a particularly elegant mathematical form. In the case of relativity, this had to await the four-dimensional geometry of Hermann Minkowski (1864--1909) and the theory's further elaboration in terms of it by Arnold Sommerfeld (1868--1951), Max Laue (1879--1960), and others \citep{Janssen and Mecklenburg 2006}.  Even so, a modern reader will have no trouble recognizing special relativity in Einstein's 1905 paper. The same reader, however, will probably only start recognizing matrix mechanics in two follow-up papers to the {\it Umdeutung} paper,  \citep{Born and Jordan 1925b} and  \citep{dreimaenner}, the famous {\it Dreim\"annerarbeit}.\footnote{During a lunch break in his AHQP interview, Alfred Land\'e (1888--1976) told Heilbron and Kuhn:  ``Heisenberg stammered something. Born made sense of it" (p.\ 10a of the transcript of sessions 1--4 of the interview; cf.\ note \ref{lande2}). Kuhn and Heilbron report that they wrote this down right after the conversation took place and call it a ``Quasi-Direct Quote."\label{lande1}} Born first recognized that Heisenberg's new non-commuting quantities are matrices. Born and Jordan first introduced the familiar commutation relations for position and momentum. In the {\it Umdeutung} paper Heisenberg had used the Thomas-Kuhn sum rule, a by-product of the Kramers dispersion formula, as his fundamental quantization condition. As we shall see, Van Vleck had actually been the first to find the sum rule, although he did not publish the result.

In the collective memory of the physics community, major discoveries understandably tend to get linked to singular events even though they are almost invariably stretched over time. The ``discovery" of the electron by J.\ J.\ Thomson (1856--1940) in 1897 or the ``discovery" of the quantum of action by Max Planck (1858--1947) in 1900 are well-known examples of this phenomenon. Special relativity is another good example of a ``discovery" that came to be associated with a single flash of insight, Einstein's recognition of the relativity of simultaneity, and  a single emblematic text,  ``On the electrodynamics of moving bodies" \citep{Einstein 1905}. Much the same can be said about Heisenberg's famous trip to Helgoland in June 1925 to seek relief from his seasonal allergies and the {\it Umdeutung} paper resulting from his epiphany on this barren island. The way in which such stories become part of physics lore can be seen as a manifestation of what Robert K.\ \citet{Merton 1968} has dubbed the ``Matthew effect," the disproportional accrual of credit to individuals perceived (sometimes retroactively) as leaders in the field.\footnote{The effect is named for the following passage from the Gospel According to St.\ Matthew: ``For unto everyone that hath shall be given, and he shall have in abundance: but from him that hath not shall be taken away even that which he hath."} 
We do, of course, recognize the singular importance of the contributions of Einstein to special relativity and of Heisenberg to matrix mechanics. But there is no need to exaggerate the extent of their achievements. They may have been the first to enter the promised land, to use another admittedly strained biblical metaphor, but they would never have laid eyes on it without some Moses-figure(s) leading the way. 

In his biography of Kramers, Dresden makes a convincing case that his subject deserves more credit for matrix mechanics than he received: ``Kramers certainly hoped and probably expected to be the single author of the Kramers-Heisenberg paper. It is probably futile to speculate how the credit for the discovery of matrix mechanics would have been distributed in that case. There would be an indispensable preliminary paper by Kramers alone, followed by a seminal paper by Heisenberg; this might well have altered the balance of recognition" \citep[p.\ 252]{Dresden}. Citing this passage, Dirk ter Haar (1998, p.\ 23), like Dresden one of Kramers' students, raises the question whether Kramers would have shared Heisenberg's 1932 Nobel prize in that case. In a review of Dresden's book, however, Nico van Kampen, another one of Kramers' students, takes issue with the pattern of  ``near misses" that \citet[pp.\ 446--461]{Dresden} sees in Kramers' career, the discovery of matrix mechanics being  one of them \citep[285--288]{Dresden}. Van Kampen asks: ``Is it necessary to explain that, once you have, with a lot of sweat and tears, constructed a dispersion formula on the basis of the correspondence principle, it is not possible to forget that background and that it takes a fresh mind to take the next step?" \citep{Van Kampen}. Similar claims can be made and similar questions can be raised in the case of Van Vleck, even though his work, unlike that of Kramers, did not directly influence Heisenberg. It did, however, by their own admission, strongly influence Born and Jordan.

Van Vleck's contribution has receded even further into the background in the history of quantum mechanics than  Kramers'. \citep{Van Vleck 1924b, Van Vleck 1924c} is not discussed in any of the currently standard secondary sources on quantum dispersion theory and matrix mechanics, such as  \citep{Jammer 1966},  \citep{Dresden}, or \citep{Darrigol}. Nor is it mentioned in Vol.\ 2 of \citep{Mehra Rechenberg} on the discovery of matrix mechanics, although it is discussed briefly in Vol.\ 1 (pp.\ 646--647) on the old quantum theory.\footnote{It is also mentioned in \citep[pp.\ 330--331]{Van der Waerden and Rechenberg 1985} and in \citep[pp.\ 131--132]{Hund 1984}. As noted in \citep[Vol.\ 6, p.\ 348, note 407]{Mehra Rechenberg}, Van Vleck's work is discussed prominently in a paper by Hiroyuki \citet{Konno 1993} on Kramers' dispersion theory.} That he worked in faraway Minnesota rather than in Copenhagen or G\"{o}ttingen, we surmise, is a major factor in this neglect of Van Vleck. Whatever the reason, the neglect is regrettable. For a modern reader, it is much easier to see in \citep{Van Vleck 1924b, Van Vleck 1924c} than in \citep{Kramers and Heisenberg 1925} or in \citep{Born 1924b} that matrix mechanics did not come as a bolt out of the blue, but was the natural outgrowth of earlier applications of the correspondence principle to the interaction of radiation and matter. 

\citet{Aitchison 2004} have recently given a detailed reconstruction of the notoriously opaque mathematics of  \citep{Heisenberg 1925c}. By way of motivating their enterprise, they quote the confession of Steven \citet{Weinberg 1992} that he has ``never understood Heisenberg's {\it motivations} for the mathematical steps in his paper" (p.\ 67; our emphasis). These authors clearly explain the mathematical steps. The motivations for these steps, however, cannot be understood, we submit, without recourse to the dispersion theory leading up to his paper. And if we want to retrace Heisenberg's steps on his sojourn to Helgoland, Van Vleck may well be our best guide.

\subsection{Structure of our paper}

Like Van Vleck's 1924 paper, our paper comes in two parts, the second  providing the technical results needed to understand the first in full detail. To provide some context for Van Vleck's work, undertaken far from the European centers in quantum theory, we begin Part One by addressing the question of America's ``coming of age" in theoretical physics in the 1920s (sec.\ 2). In sec.\ 3, we relate the story of how matrix mechanics grew out of dispersion theory in the old quantum theory, drawing on the extensive secondary literature on this episode as well as on the materials brought together in the AHQP. This story is usually told from a Eurocentric perspective. Following our discussion in sec.\ 2, we shall look at it from a more American vantage point. Discussion of the famous BKS theory \citep{BKS}, which is prominently mentioned in many papers on dispersion theory in 1924--1925, is postponed until  sec.\ 4. In this context we shall pay special attention to the role of Van Vleck's fellow graduate student at Harvard, John C.\ Slater (1900--1976).\footnote{On Slater, see, e.g., \citep{Schweber 1990}.} The reason for keeping the discussion of BKS separate from the discussion of dispersion theory is that we want to argue that the rise and fall of BKS was largely a sideshow distracting from the main plot line, which runs directly from dispersion theory to matrix mechanics. In hindsight, BKS mainly deserves credit for the broad dissemination of its concept of `virtual oscillators.' Contrary to widespread opinion, both among contemporaries and among later historians, these virtual oscillators did not originate in the BKS theory. They were introduced the year before, under a different name and in the context of dispersion theory, by the Breslau physicists Rudolf Ladenburg (1882--1952) and Fritz Reiche (1883--1969), who called them `substitute oscilators'  [{\it Ersatz\-oszillatoren}\footnote{We follow the translation used in \citep[e.g., 139]{Konno 1993}.}]  \citep[p. 588, p. 590]{Ladenburg and Reiche 1923}. This paper is important in its own right and underscores the  key achievement of Van Vleck's two-part paper. Both \citet{Van Vleck 1924b, Van Vleck 1924c} and \citet{Ladenburg and Reiche 1923} used the correspondence principle to construct quantum expressions for emission, absorption, and dispersion. Van Vleck provides impeccable constructions of all three; Ladenburg and Reiche made serious errors in the case of both dispersion and absorption. The expertise Van Vleck had gained in classical mechanics by working on the problem of helium in the old quantum theory \citep{Van Vleck 1922a, Van Vleck 1922b} 
put him in an ideal position to correct these errors. We suggest that, at least in part, he may have wanted to do just that with \citep{Van Vleck 1924b, Van Vleck 1924c}.\footnote{\citep{Ladenburg 1921} and \citep{Ladenburg and Reiche 1923} are cited in \citep[p.\ 339]{Van Vleck 1924b}.}

In sec.\ 5, the first section of Part Two, we give an elementary and self-contained presentation, following  \citep{Van Vleck 1924b, Van Vleck 1924c}, of the technical results on which our narrative in secs.\ 3 and 4 rests. In particular, we use canonical perturbation theory in action-angle variables to derive a classical formula for the dispersion of radiation by a charged harmonic oscillator and apply the correspondence principle to that formula to obtain the famous Kramers dispersion formula for this special case. This fills an important pedagogical gap in the historical literature. Given the central importance of the Kramers dispersion formula for the development of quantum mechanics, it is to be lamented that there is no explicit easy-to-follow derivation of this result in the extensive literature on the subject. In the later parts of sec.\ 5 and in sec.\ 6, we take a closer look at Van Vleck's main concerns in his 1924 paper, which was absorption rather than dispersion and the extension of results for the special case of a charged harmonic oscillator (which suffices to understand how matrix mechanics grew out of dispersion theory) to  arbitrary non-degenerate multiply-periodic systems. In sec.\ 7, we present a simple modern derivation of the Kramers dispersion formula and related results, which we hope will throw further light on derivations and results in secs.\ 5 and 6 as well as on the narrative in secs.\ 3 and 4. Finally, in sec.\ 8, we bring together the main conclusions of our investigation.

\section{Americans and quantum theory in the early 1920s}
``[A]lthough we did not start the orgy of quantum mechanics, our young theorists joined it promptly" \citep[p.\ 24]{Van Vleck 1964}.\footnote{Quoted and discussed in \citep[p.\ 456]{Coben 1971}} This is how our main protagonist, known to his colleagues simply as `Van',
described the American participation in the quantum revolution of the mid-1920s for an audience in Cleveland in 1963. Van Vleck spoke as the first recipient of an award named for America's first Nobel prize winner in physics, Albert A.\ Michelson (1825--1931). He was fond of the ``orgy"-metaphor, which he had picked up from the German-American physicist Ralph Kronig (1904--1995). In his Michelson address he mentioned how he had used it off-handedly a few years earlier during a press conference at Harvard on Russian contributions to science, only to find himself quoted in a newspaper as saying that there had been a ``Russian orgy in theoretical physics" (ibid.). He was selling himself and his countrymen short, however, by characterizing the American contribution to the quantum revolution as simply a matter of  joining an orgy started by the Europeans and in full swing by the time the Americans arrived on the scene. 

Eight years later, Van Vleck, in fact, took exception to what sounds like a similar characterization given by another leading American physicist of his generation, Isidor I.\ Rabi (1898--1988). Van Vleck quoted a comment that Rabi made in a TV documentary about Enrico Fermi (1901--1954): 
\begin{quotation}
We had produced a large number of people who had been brought up to a certain level, then needed some help, some leadership to get over the hump. Once they were over the hump they were tremendous. People of my generation brought them over the hump, largely from attitudes, tastes, and developments which we had learned in Europe \citep[p.\ 7]{Van Vleck 1971}.
\end{quotation}
As Kuhn and others have emphasized, Rabi's point was that American physicists returning from Europe rather than European \'emigr\'es were mainly responsible for the coming of age of American physics.\footnote{\label{Uhlenbeck}See p.\ 20 of the transcript of the last of five sessions of Kuhn's AHQP interview with George E.\ Uhlenbeck (1900--1988).} This issue has been hotly debated in the history of physics  literature.\footnote{For a concise summary and detailed references to the older literature, see  \citep[pp.\ 171--173]{Moyer}. Whereas our focus will be on American contributions to atomic physics, \citet{Assmus 1992b, Assmus 1999} has argued that American theoretical physics came of age in molecular physics (cf.\ note \ref{assmuss} below).} Our study of some early American contributions to quantum theory supports the observation  by Sam 
\citet{Schweber 1986} that in the 1930s theoretical physics was ``already a thriving enterprise in the United States. The refugee scientists resonated with and reinforced American strength and methods: they did not create them" (p.\ 58).

Commenting on Rabi's remark, \citet{Van Vleck 1971} reiterated the point of his Michelson address that ``quantum mechanics was a basically European discovery" (p.\ 6). In 1928, he had likewise characterized it as ``the result of the reaction of mind on mind among European talent in theoretical physics" \citep[p.\ 467]{Van Vleck 1928a}. In 1971, however, he added that ``there has been too much of an impression that American physicists, even in the application of quantum mechanics, were effective only because they had the aid of European physicists, either by going to Europe, or because of their migration to America" \citep[p.\ 6]{Van Vleck 1971}. Van Vleck, who was proud to be a tenth-generation American,\footnote{He could trace his ancestry back to the fifteenth century, to a certain Johan van Vleeck of Maastricht. One of the latter's descendants, Tielman van Vleeck (or von Fleck), left Bremen for New Amsterdam in 1658  \citep[pp.\ 5--6]{Fellows}} received his entire education in the United States. He hardly had any contact with European physicists before 1925, although he did meet a few on a trip to Europe with his parents in the summer of 1923.  In Copenhagen, he called on Bohr, who suggested that he get in touch with Kramers,\footnote{See p.\ 14 of the transcript of session 1 of the AHQP interview with Van Vleck.} Bohr's right-hand man throughout the period of interest to us. Kramers was not in Denmark at the time but in his native Holland. Decades later, when he received the prestigious Lorentz medal from the {\it Koninklijke Akademie van Wetenschappen} in Amsterdam, Van Vleck recalled how he had searched for Kramers high and low.  After he had finally tracked him down---it can no longer be established whether this was in Bergen aan Zee or in Schoorl---the two men went for a long walk in the dunes along the North-Sea coast: ``This was the beginning of a friendship that lasted until his passing in 1952" \citep[p.\ 9]{Van Vleck 1974}. Unfortunately, Van Vleck does not tell us what he and Kramers talked about. 

\subsection{Education}
 
Van Vleck learned the old quantum theory of Bohr and Sommerfeld at Harvard as one of the first students to take the new course on quantum theory offered by  Edwin  C.\ Kemble (1889--1984), the first American physicist to write a predominantly theoretical quantum dissertation. Kemble, \citet{Van Vleck 1992} wrote in a biographical note accompanying the published version of his Nobel lecture,  ``was the one person in America at that time qualified to direct purely theoretical research in quantum atomic physics" (p.\ 351). Kemble's course roughly followed \citep{Sommerfeld 1919}, the bible of the old quantum theory. Van Vleck supplemented his studies by reading \citep{Bohr 1918} and \citep{Kramers 1919} \citep[p.\ 17]{Fellows}. 

Van Vleck was part of a remarkable cohort of young American quantum theorists, which also included Slater, Gregory Breit (1899--1981), Harold C.\ Urey (1893--1981), and David M.\ Dennison (1900--1976).
Just as Van Vleck was the first to write a purely theoretical dissertation at Harvard in 1922, Dennison was the first to do so at the University of Michigan in 1924.\footnote{\label{dennison}See p.\ 10 of the transcript of the first of three sessions of Kuhn's AHQP interview with Dennison.} Dennison could take advantage of the presence of Oskar Klein (1894--1977), an early associate of Bohr,\footnote{See \citep{O. Klein 1967} for his reminiscences about his early days in Copenhagen.} who was a visiting faculty member in the physics department in Michigan from 1923 to 1925 \citep[p.\ 321]{Sopka}. This is where Klein came up with what is now known as the Klein-Gordon equation; it is also where he made his contribution to what is now known as the Kaluza-Klein theory.\footnote{See p.\ 13 of the transcript of session 5 of the AHQP interview with Uhlenbeck.} 

Reminiscences about the early days of quantum physics in the United States can be found in \citep{Van Vleck 1964,Van Vleck 1971} and in \citep{Slater 1968,Slater 1973,Slater 1975}. It is also an important topic of conversation in the AHQP interviews with Van Vleck, Slater, Dennison, and Kemble in the early 1960s. These interviews need to be handled with care. In the case of Slater and Van Vleck, one can say, roughly speaking, that the former had a tendency to exaggerate the importance of American contributions, especially his own, while the latter tended to downplay  their importance. In sharp contrast,  for instance, to the modest remarks by Van Vleck quoted in sec.\ 1.1, Slater boasted that he ``was really working toward quantum mechanics before quantum mechanics came out. I'm sure if it was delayed a year or so more, I would have got it before the others did."\footnote{P.\ 40 of the transcript of the first session of the AHQP interview with Slater.}

The older generation---men such as Michelson and Robert A.\ Millikan (1868--1953)---recognized that the United States badly needed to catch up with Europe in quantum physics. The Americans were already doing first-rate experimental work.  One need only think of \citet{Millikan 1916} verifying the formula for the photo-electric effect \citep{Stuewer forthcoming}
or of Arthur H.\ Compton (1892--1962) (1923) producing strong evidence for the underlying hypothesis of light quanta \citep{Stuewer 1975}. Theory, however, was seriously lagging behind. In 1963, Van Vleck's teacher, Ted Kemble, recalled:
\begin{quotation}
[T]he only theoretical physicists in the country at that time were really men on whom the load of teaching all the mathematical physics courses lay, and they all spent their time teaching. It wasn't, as I remember, a constructive occupation.\footnote{P.\ 4 of the transcript of the last two of three sessions of the AHQP interview with Kemble. See also p.\ 10 of the transcript of the first of session.}
\end{quotation} 
The one theorist who, in Kemble's estimation, was active in research in classical physics, Arthur Gordon Webster (1863--1923), was never able to make the transition to quantum theory. Webster, Kemble said,
\begin{quotation}
just couldn't keep up with what was going on when the quantum theory began. I always understood that the reason he killed himself was simply because he discovered that suddenly physics had gone off in a new direction and he was unable to follow, and couldn't bear to take a seat in the back and be silent.\footnote{P.\ 12 of the transcript of the first session of the AHQP interview with Kemble.}
\end{quotation} 
When quantum theory arrived on the scene, some experimentalists tried their hands at teaching it themselves \citep[p.\ 444]{Coben 1971}. In this climate, young American physicists with a knack for theory became a hot commodity. They received fellowships to learn the theory at the feet of the masters in Europe and offers of faculty positions straight out of graduate school.\footnote{For further discussion of quantum physics in America before the mid-30s, see \citep{Coben 1971}, \citep{Seidel 1978}, \citep[pp.\ 168--169]{Kevles 1978},  \citep{Weart 1979}, \citep{Schweber 1986}, \citep{Holton 1988}, and, especially, \citep{Sopka}.} 

\subsection{Postdocs and faculty positions}

The careers of  the young theorists listed above amply illustrate the new opportunities in the mid-1920s. Slater went to Europe on a Sheldon fellowship from Harvard and spent the first half of 1924 with Bohr and Kramers in Copenhagen. During this period, Urey and Frank C.\ Hoyt (1898--1977) were in Copenhagen as well, Urey on a small fellowship from the American-Scandinavian Foundation, Hoyt on a more generous NRC fellowship paid for by the Rockefeller foundation.\footnote{See \citep[p.\ 157]{Robertson 1979}, \citep[pp.\ 71, 97]{Sopka}, and Slater to Van Vleck, July 27, 1924 (AHQP) \label{letter 1}} Among the visitors the Americans got to meet in Bohr's institute were Heisenberg  and Pauli.
Hoyt, a promising student who never reached the level of distinction of the cohort immediately following him,\footnote{He wrote several papers on applications of Bohr's correspondence principle \citep{Hoyt 1923, Hoyt 1924, Hoyt 1925a, Hoyt 1925b}. The first two are cited in \citep[p.\ 334]{Van Vleck 1924b} and all but the second are cited in \citep[pp.\ 124, 146]{Van Vleck 1926a}. \citep{Ladenburg and Reiche 1924} cites the second paper, referring to the author as ``W.\ C.\ Hoyt" (p.\ 672).  Hoyt also translated Bohr's Nobel lecture into English \citep{Bohr 1923a}. Hoyt ended up making a career in weapons research rather than in academic physics. After the war, he worked at Argonne National Laboratory, Los Alamos, and Lockheed. He was  interviewed for the AHQP by Heilbron but did not remember much of the early days of quantum theory.} was in Copenhagen for almost two years, from October 1922 to September 1924, Urey for less than one, from September 1923 to June 1924, and Slater only for a few months, from December 1923 to April 1924. Slater did not have a good experience in Copenhagen. This transpires, for instance, in the letter he wrote to Van Vleck on his way back to the United States. Off the coast of Nantucket, a few hours before his ship---The Cunard R.M.S.\ ``Lancastria"---docked in New York, he wrote:
\begin{quotation}
Don't remember just how much I told you about my stay in Copenhagen. The
paper with Bohr and Kramers  [proposing the BKS theory] was got out of the way the first six weeks or
so---written entirely by Bohr and Kramers.  That was very nearly the only
paper that came from the institute at all the time I was there; there
seemed to be very little doing. Bohr does very little and is chronically
overworked by it \ldots Bohr had to go
on several vacations in the spring, and came back worse from each one.\footnote{Slater to Van Vleck, July 27, 1924 (AHQP). The second sentence of this passage is quoted by \citet[p.\ 165]{Dresden} in the course of his detailed discussion of Slater's reaction to his experiences in Copenhagen. }
\end{quotation}
In October 1924, Dennison arrived in Copenhagen, on an {\it International Education Board} (IEB) fellowship, another fellowship paid for by the Rockefeller Foundation.\footnote{Bohr arranged for one of these fellowships to pay for Heisenberg's visit to Copenhagen in the fall of 1924 \citep[pp.\ 180, 183]{Cassidy}.
See also the acknowledgment in \citep[p.\ 860]{Heisenberg 1925b}.} The state of quantum theory in America was already beginning to change at that point. Like Hoyt, Dennison was awarded a NRC fellowship, but was told that he could only spend the money at an American institution.\footnote{See p.\ 12 of the transcript of session 1 of the AHQP interview with Dennison. In 1923, the NRC had likewise rejected the proposal of Robert S.\ Mulliken (1896--1987) to go work with Ernest Rutherford (1871--1937) in Cambridge. Mulliken became a NRC research fellow at Harvard instead \citep[p.\ 23]{Assmus 1992b}.}

Van Vleck and Slater, who both started graduate school at Harvard in 1920 (Van Vleck in February, Slater in September) and lived in the same dormitory,\footnote{See Van Vleck, {\it 1920--1930. The first ten years of John Slater's scientific career.} Unpublished manuscript, American Institute of Physics (AIP), p.\ 2.} had at one point discussed going to Copenhagen together upon completion of their Ph.D.\ degrees in 1923. In the end, Van Vleck went to Minneapolis instead. In the biographical note accompanying his Nobel lecture from which we already quoted above, he reflected:
\begin{quotation}
I was fortunate in being offered an assistant professorship at the University of Minnesota 
\ldots with purely graduate courses to teach. This was an unusual move by that institution, as at that time, posts with this type of teaching were usually reserved for older men, and recent Ph.D.'s were traditionally handicapped by heavy loads of undergraduate teaching which left little time to think about research \citep[p.\ 351]{Van Vleck 1992}.
\end{quotation}
When the university hired Van Vleck it also hired Breit so that its new recruits would not feel isolated.\footnote{See p.\ 14 and p.\ 18 of the transcript of session 1 of the AHQP interview with Van Vleck.} 
Breit is one of the more eccentric figures of 20th-century American physics. He was born in Russia and came to the United States in 1915. In a biographical memoir of the {\it National Academy of Sciences}, we read that 
\begin{quotation}
John Wheeler relates a story told to him by Lubov [Gregory's sister] that she and Gregory were vacationing on the sea when the call to leave Russia came, and they `came as they were.' For Gregory this meant dressed in a sailor suit with short pants; he was still wearing it when he enrolled in Johns Hopkins (at age sixteen!). Wheeler attributes some of Gregory's subsequent reticence to the ragging he took at the hand of his classmates for his dress \citep[pp.\ 29--30]{Hull 1998}. 
\end{quotation}
In the run-up to the Manhattan Project, Breit served as ``Coordinator of Rapid Rupture." He was obsessed with secrecy and resigned in May 1942. He was replaced by J.\ Robert Oppenheimer (1904--1967) \cite[p.\ 48]{Goodchild}. True to form, Breit declined to be interviewed for the AHQP. In a memorandum dated April 8, 1964 (included in the folder on Breit in the AHQP), Kuhn describes how they met for lunch, but did not get beyond ``casual reminiscences." Kuhn ends on a positively irritated note: ``we broke off amicably but with zero achievement to report for the project." 

Breit and Van Vleck replaced W.\ F.\ G.\ Swann (1884--1962) who had left Minneapolis for Chicago, taking his star graduate student Ernest O.\ Lawrence (1901--1958) with him. As \citet{Van Vleck 1971} notes wryly: ``A common unwitting remark of the lady next to me at a dinner party was ``Wasn't it too bad Minnesota lost Swann---it took two men to replace him!"" (p.\ 6). 

Just as Minnesota hired both Breit and Van Vleck in 1923,  the University of Michigan  hired not one but two  students of Paul Ehrenfest (1880--1933) in 1927, Uhlenbeck and Samuel A.\ Goudsmit (1902--1978) \citep[p.\ 460]{Coben 1971}.\footnote{See also  \citep[p.\ 149]{Sopka} and the AHQP interview with Dennison. The recruiting was done by Walter F.\ Colby (1880--1970) and Harrison M.\ Randall (1870--1969). \label{colby}} In addition Michigan hired Dennison, its own alumnus, upon his return from Copenhagen. Ann Arbor thus became an important center for quantum theory, especially in molecular physics \citep[pp.\ 4, 26, 30]{Assmus 1992b}. 
While Uhlenbeck and Goudsmit essentially remained in Ann Arbor for the rest of their careers, neither Breit nor Van Vleck stayed long in Minneapolis. Breit left for the Carnegie Institution of Washington after only one year, Van Vleck for the University of Wisconsin, his {\it alma mater}, after five. Van Vleck agonized over the decision to leave Minnesota, where he had been promoted to associate professor in June 1926 and, only a year later, to full professor \citep[Ch.\ VII]{Fellows}. Moreover, on June 10, 1927, he had married Abigail Pearson (1900--1989), whom he had met while she was an undergraduate at the University of Minnesota and who had strong ties to Minneapolis.\footnote{After her husband's death, Abigail made a generous donation to  the University of Minnesota to support the Abigail and John van Vleck  Lecture Series. Phil Anderson gave the inaugural lecture in 1983 and the series has brought several Nobel prize winners to Minneapolis since. The main auditorium in the building currently housing the University of Minnesota physics department is also named after the couple.}

To replace Van Vleck, Minnesota made the irresistible offer of a full professorship to the young Edward U.\ Condon (1902--1974).  Minnesota had offered Condon an assistant professorship the year before. At that point, Condon had received six such offers and had decided on Princeton \citep[p.\ 321]{Condon 1973}. His laconic response to this embarrassment of riches: ``The market conditions for young theoretical physicists continues [sic] to surprise me"  \citep[p.\ 463]{Coben 1971}. Before his first Minnesota winter as a full professor, Condon already regretted leaving New Jersey. He returned to Princeton the following year. Condon, Rabi, and Oppenheimer\footnote{Oppenheimer enrolled as an undergraduate at Harvard in 1922, two years after Van Vleck and Slater started graduate school there.} were the leaders of the cohort  of American quantum theorists graduating right after the quantum revolution of 1925. The cohort most relevant to our story graduated  right before that watershed.

\subsection{The Physical Review}

It was during Van Vleck's tenure in Minnesota that his senior colleague  John T. (Jack) Tate (1889--1950) took over as editor-in-chief of  {\it The Physical Review} \citep[pp.\ 142--145, 203, note 11]{Sopka}. Tate edited the journal from 1926 to 1950.\footnote{It is largely in recognition of this achievement that the current Minnesota physics building is named after him.} \citet{Van Vleck 1971} stressed the importance of Tate taking over the journal, describing it as ``another revolution" in the ``middle of the quantum revolution" (pp.\ 7--8). Van Vleck was highly appreciative of Tate's role: ``He published my papers very promptly, and also often let me see manuscripts of submitted papers, usually to referee" (ibid.). Thanks in no small measure to Van Vleck and other young whippersnappers  in quantum theory, Tate turned what had been a lack-luster publication into the prestigious journal it still is today. Van Vleck recalled the transformation:
\begin{quotation}
{\it The Physical Review} was only so-so, especially in theory, and in 1922 I was greatly pleased that my doctor's thesis [Van Vleck, 1922] was accepted for publication by the {\it Philosophical Magazine} in England  \ldots By 1930 or so, the relative standings of {\it The Physical Review} and {\it Philosophical Magazine} were interchanged \ldots Prompt publication, beginning in 1929, of ``Letters to the Editor" in {\it The Physical Review} \ldots obviated the necessity of sending notes to {\it Nature}, a practice previously followed by our more eager colleagues [see, e.g., \citep{Breit 1924b}, \citep{Slater 1924, Slater 1925c}]  \citep[pp.\ 22, 24]{Van Vleck 1964}.
\end{quotation}
Van Vleck's impression is corroborated by two foreign-born theorists who made their careers in the United States, Rabi and Uhlenbeck \citep[p.\ 456]{Coben 1971}. Rabi was born in Galicia
but moved to New York City as an infant. Rabi liked to tell the story of how, when he returned to Europe to study quantum theory in Germany in 1927, he discovered that  {\it The Physical Review} ``was so lowly regarded that the University of G\"{o}ttingen waited until the end of the year and ordered all twelve monthly issues at once to save postage" (ibid.). On other occasions, Rabi told this story about Hamburg University \citep[p.\ 4]{Rigden 1987}.  He told Jeremy \citet{Bernstein} that ``in Hamburg so little was thought of the journal \ldots that the librarian uncrated the issues only once a year" (p.\ 28). The following exchange between Kuhn and Heisenberg, talking about the early twenties, is also revealing: 
\begin{quotation}
\begin{description}
\item[] Heisenberg: ``What was the American paper at that time?"  
\item[]  Kuhn: ``The {\it Physical Review}?" 
\item[]  Heisenberg: ``No, that didn't exist at that time. I don't think so. Well, in these early times it probably didn't play a very important role."\footnote{P.\ 5 of the transcript of session 3 of the AHQP interview with Heisenberg.}
\end{description}
\end{quotation}  
In a talk about Condon, Rabi elaborated on the mediocrity of {\it The Physical Review}:
\begin{quotation}
it was not a very exciting journal even though I published my dissertation in it. And we felt this very keenly. Here was the United States, a vast and rich country but on a rather less than modest level in its contribution to physics, at least per capita. And we resolved that we would change the situation. And I think we did. By 1937 the {\it Physical Review} was a leading journal in the world  \citep[p.\ 7]{Rabi 1975}
\end{quotation}
Uhlenbeck remembered how as a student in Leyden he viewed  {\it The Physical Review} as ``one of the funny journals just like the Japanese."\footnote{See p.\ 20 of the transcript of session 5 of the AHQP interview with Uhlenbeck.} His initial reaction to the job offer from Michigan suggests that, at least at the time, his disdain for American physics journals extended to the country as a whole: ``If it had been Egypt or somewhere like that, I would have gone right away, or China, or even India, I always wanted to go to exotic places [Uhlenbeck was born in Batavia in the Dutch East Indies, now Jakarta, Indonesia]; but America seemed terribly dull and uninteresting"\citep[p.\ 460]{Coben 1971}. In the AHQP interview with Uhlenbeck, one finds no such disparaging comments. In fact, Uhlenbeck talks about how he had reluctantly agreed to return to the Netherlands in 1935 to replace Kramers, who had left Utrecht for Leyden to become Ehrenfest's successor after the latter's suicide.\footnote{See p.\ 9 of the transcript of session 5 of the AHQP interview with Uhlenbeck..} Uhlenbeck was back in Ann Arbor in 1939.

\subsection{The lack of recognition of early American contributions to quantum theory}

Given the disadvantage they started out with, American theorists in the early 1920s would have done well had they just absorbed the work of their European counterparts and transmitted it to the next generation. They did considerably better than that. Even before the breakthrough of \citet{Heisenberg 1925c} they started making important contributions themselves. According to Alexi \citet{Assmus 1992b}, ``[a]tomic physics was shark infested waters and was to be avoided; U.S.\ physicists would flourish and mature in the calmer and safer tidepools of molecular physics" (p.\ 8; see also Assmus, 1999, p.\ 187). She sees the early contributions of Van Vleck and Slater to atomic physics, which will be the focus of our study, as exceptions to this rule:
\begin{quotation}
Van Vleck and Slater viewed themselves as the younger generation, as central figures in the ``coming of age" of U.S.\ physics. They had been given the knowledge that Kemble and his generation could provide and felt themselves capable of pushing into areas where the physics community in the United States had not dared to venture. Still, after experiences had muted their youthful exuberance, they turned to the by-then traditional problems of American quantum physics[,] problems that addressed the building up of matter rather than its deconstruction \citep[p.\ 22]{Assmus 1992b}. 
\end{quotation}
We hope to show that American work in atomic physics was significantly more important---if not in quantity, then at least in quality---than these remarks suggest.\footnote{Assmus is probably right, however, that the Americans contributed more to molecular than to atomic physics. This would fit with the thesis of \citet{Schweber 1990} that ``Americans contributed most significantly to the development of quantum mechanics in quantum chemistry" (pp.\ 398--406)\label{assmuss}} 
Slater was one of the architects of the short-lived but highly influential Bohr-Kramers-Slater (BKS) theory \citep{BKS} (see sec.\ 4). 
Van Vleck's two-part article in {\it The Physical Review} \citep{Van Vleck 1924b, Van Vleck 1924c}  is less well-known. 

Originally, Van Vleck's paper was to have three parts. A rough draft of the third part has been preserved.\footnote{{\it American Institute of Physics}, Van Vleck papers, Box 17. We are grateful to Fred Fellows for sharing a copy of this manuscript with us.} Van Vleck did not finish the third part at the time. As he explained in a letter to Born on November 30, 1924 (AHQP): ``Part III  which is not yet ready relates to classical black body radiation rather than quantum theory." It was only toward the end of his life that he returned to the masterpiece of his youth. Three years before he died he published a paper, co-authored with D.\ L.\ Huber, that can be seen as a substitute for part III. As the authors explain:
\begin{quotation}
Part III was to be concerned with the equilibrium between absorption and emission under the Rayleigh-Jeans law. It was never written up for publication because in 1925 the author was busy writing his book [Van Vleck, 1926a] and of course the advent of quantum mechanics presented innumerable research problems more timely than a purely classical investigation. The idea occurred to him to use the 50th anniversary of Parts I and II as the date for publishing a paper which would start with Part III and might even bear its title. Although he did not succeed in meeting the deadline, it still provided a partial motivation for collaborating on the present article \citep[p.\ 939]{Van Vleck and Huber 1977}. 
\end{quotation}
It was at the suggestion of Jordan, that  van der Waerden included the first (quantum) part of Van Vleck's 1924 paper in his anthology on matrix mechanics \citep[see the preface]{Van der Waerden}.\footnote{See also \citep[pp.\ 110--111]{Sopka}.} 
In his interview with Van Vleck for the AHQP in October 1963, Kuhn claimed that Jordan had told him that Born and Jordan  ``were working quite hard in an attempt to reformulate it [Van Vleck, 1924b,c] and had been multiplying Fourier coefficients together,\footnote{The multiplication of quantum-theoretical quantities corresponding to classical Fourier components is one of the key elements of Heisenberg's {\it Umdeutung} paper.} just at the time they got the Heisenberg paper that was going to be matrix mechanics."\footnote{See p.\  24 of the transcript of session 1 of the AHQP interview with Van Vleck.} 
In fact, a paper by \citet{Born and Jordan 1925a} building on \citep{Van Vleck 1924b, Van Vleck 1924c} was submitted to {\it Zeitschrift f\"ur Physik} on June 11, 1925, several weeks {\it before} Heisenberg's breakthrough  \citep[p.\ 198]{Cassidy}.  We therefore suspect that Kuhn just misremembered or misconstrued what Jordan had told him during an interview for the AHQP in June 1963, a few months before the interview with Van Vleck. Van Vleck's paper is discussed prominently in the interview with Jordan. It is first brought up in the second session (see p.\ 14 of the transcript). In this exchange Kuhn insisted that \citep{Born and Jordan 1925a} had come out {\it before} \citep{Van Vleck 1924b, Van Vleck 1924c}. Jordan corrected Kuhn at the beginning of the third session, which prompted some further discussion of Van Vleck's paper. However, it was Kuhn, not Jordan, who suggested at that point that Born and Jordan continued to pursue the ideas in Van Vleck's paper even after publishing  \citep{Born and Jordan 1925a}. Jordan did not confirm this. Still, although Kuhn probably embellished he story, there is no question that Van Vleck's paper had a big impact on the work of Born and Jordan. Jordan emphasized this in the interview with Kuhn, in a letter to van der Waerden of December 1, 1961 (quoted in Van der Waerden, 1968, p.\ 17), and in \citep{Jordan 1973}. We quote from this last source:
\begin{quotation}
Van Vleck gave a derivation of Einstein's laws of the relation between the probabilities of spontaneous emission and positive and negative absorption. This result of Einstein's had been looked upon  for a long time in a sceptical manner by Niels Bohr; now it was highly interesting to see, just how from Bohr's preferred way of thinking, a derivation of Einstein's law could be given. Born and I performed a simplified mathematical derivation of the results of Van Vleck. Our article on this topic [Born and Jordan, 1925a] {\it did not contain anything new apart from our simpler form of the calculation}, but by studying this topic we both came to a more intimate understanding of Bohr's leading ideas \citep[p.\ 294, our emphasis]{Jordan 1973}.\footnote{See secs.\ 5.2, 5.3 and 6.1 below for discussion of Van Vleck's correspondence principles for emission and absorption. As in the case of \citep{Kramers and Heisenberg 1925}, we suspect that \citep{Born and Jordan 1925a} is actually more difficult to follow for most modern readers than \citep{Van Vleck 1924b, Van Vleck 1924c}.}
\end{quotation}

Incidentally, \citet[p.\ 7]{Van Vleck 1971} pointed to  this important pre-1925 contribution of his own as well as to Slater's role in BKS and Kemble's work on helium to demonstrate the inaccuracy of Rabi's characterization of American work in quantum theory quoted earlier. Even at the time, Van Vleck had felt that the Europeans were not giving the Americans their due. He complained about this in a letter to Born:
\begin{quotation}
I am writing this letter regarding some of the references to my work in your articles.  I fully realize that an occasional error in a reference is unavoidable, for I have made such mistakes myself. I would gladly overlook any one error, but inasmuch as there are two or three instances, it is perhaps worth while to call them to your attention.
On p.\ 332 of your treatise on ``Atommechanik" [Born, 1925],  the reference to my work on the crossed-orbit model of the normal helium atom is given as [Van Vleck, 1923].  This reference is only to the abstract of some work on {\it excited} helium and the references to my articles  on {\it normal helium} are [Van Vleck, 1922a] \ldots
and especially [Van Vleck, 1922b],  where the details of the computations are given.  This incorrect reference to a paper on another subject published a year later makes it appear as though my computation was published simultaneously or later than that of Kramer[s] [\citep{Kramers 1923}, cited in the same footnote as \citep{Van Vleck 1923} in \citep[p.\ 332]{Born 1925}].   The same error is also found in your article [Born, 1924b] on perturbation theory \ldots  Also in your book on Atommechanik [\citep[p.\ 332]{Born 1925}, the sentence with the footnote referring to \citep{Kramers 1923} and \citep{Van Vleck 1923}] you say ``das raumliche [sic] Modell ist ebenfalls von Bohr vorgeschlagen" [the spatial model has also been proposed by Bohr], without any mention of the name Kemble, who proposed the crossed-orbit model in [Kemble, 1921] before [Bohr, 1922].\footnote{Van Vleck to Born, October 19, 1925, draft (AHQP).}
\end{quotation}
Van Vleck then comes to the most egregious case, Born's failure to properly acknowledge his two-part paper on the correspondence principle in \citep{Born and Jordan 1925a}. Especially in view of Jordan's comments on the importance of this paper quoted above, the authors were very stingy in giving him credit. 
Van Vleck's letter continues:
\begin{quotation}
I was much interested in your recent article on the Quantization of Aperiodic Systems,  in which you show that the method of Fourier integrals gives many results obtained by ``Niessen and Van Vleck" [Born and Jordan, 1925a, p.\ 486], placing my name after Niessen's [Kare Frederick Niessen (1895--1967)], even though his paper [Niessen, 1924] did not appear until Dec.\ 1924   while the details of my computations were given in the Physical Review for Oct.\ 1924 [Van Vleck, 1924b, 1924c] and a preliminary notice published in the Journal of the Optical Society for July 1924 [Van Vleck, 1924a],  before Niessen's article was even submitted for publication.  I think you wrote me inquiring about my work shortly after the appearance of this preliminary note, and so you must be aware that it was the first to appear  \ldots  inasmuch as Niessen's discussion is somewhat less general than my own, it seems to me that it scarcely merits being listed first (Ibid.).
\end{quotation}
Writing from Cambridge, Massachusetts, where he was visiting MIT, Born apologized.\footnote{Born to Van Vleck, November 25, 1925 (AHQP). Born had been less generous in the case of a similar complaint from America a few years earlier (see sec.\ 3 below).} 
Born had indeed written to Van Vleck concerning \citep{Van Vleck 1924a}, albeit a little later than the latter remembered:
\begin{quotation}
While we already came close to one another in the calculation of the helium atom, I see from your paper ``A Correspondence Principle for Absorption" [Van Vleck, 1924a] that we now approach each other very closely with our trains of thought \ldots I am sending you my paper ``On Quantum Mechanics" [Born, 1924], which pursues a goal similar to yours.\footnote{Born to Van Vleck, October 24, 1924 (AHQP).}
\end{quotation}
This goes to show---Rabi's anecdotal evidence to the contrary notwithstanding---that at least some European physicists did keep up with theoretical work published in American journals,  the {\it Journal of the Optical Society of America} in this case,  even if they were not particularly generous acknowledging its importance in print.

\section{Dispersion theory as the bridge between the old quantum theory and matrix mechanics} 

From the point of view of modern quantum mechanics, the old quantum theory of Bohr and Sommerfeld---especially in the hands of the latter and members of his Munich school---was largely an elaborate attempt at damage control. In classical physics the state of a physical system is represented by a point in the phase space spanned by a system's generalized coordinates  and momenta $(q_i, p_i)$. All its properties are represented by functions $f(q_i, p_i)$ defined on this phase space. In quantum mechanics the state of a system is represented by a ray in the Hilbert space associated with the system; its properties are represented by operators acting in this Hilbert space, i.e., by rules for {\it transitions} from one ray to another. In the old quantum theory, one bent over backward to retain classical phase space. Quantum conditions formulated in various ways in  \citep{Sommerfeld 1915a}, \citep{Wilson 1915}, \citep{Ishiwara 1915}, \citep{Schwarzschild 1916}, and \citep{Epstein 1916} only restricted the allowed orbits of points in phase space. These conditions restricted the value of so-called action integrals for every degree of freedom of some multiply-periodic system to integer multiples of Planck's constant $h$,
\begin{equation}
\oint p_i dq_i = n_i h,
\label{eq:1.0}
\end{equation}
where the integral is extended over one period of the generalized coordinate $q_i$ (there is no summation over $i$).

Imposing such quantum conditions on classical phase space would not do in the end. As the picture of the interaction of matter and radiation in the old quantum theory already suggests, more drastic steps were required. In Bohr's theory the frequency $\nu_{i\rightarrow f}$ of the radiation emitted when an electron makes the transition from an initial state $i$  to a final state $f$  is given by the energy difference $E_{i}-E_{f}$ between the two states divided by $h$. Except in the limiting case of high quantum numbers, this radiation frequency differs sharply from the frequencies with which the electron traverses its quantized orbits in classical phase space before and after emission. This was widely recognized as the most radical aspect  of the Bohr model.  Erwin Schr\"odinger (1887--1961), for instance, opined in 1926 that this discrepancy between radiation frequency and orbital frequency
 \begin{quotation}
\ldots seems to me, (and has indeed seemed to me since 1914), to be something so {\it monstrous}, that I should like to characterize the excitation of light in this way as really almost {\it inconceivable}.\footnote{Schr\"odinger to Lorentz, June 6, 1926 \citep[p.\ 61]{M. Klein 1967}.}
\end{quotation} 
Imre \citet{Lakatos} produces a lengthy quotation from an obituary of Planck by \citet{Born 1948}, in which the same point is made more forcefully. It even repeats some of the language of Schr\"{o}dinger's letter:
\begin{quotation}
That within the atom certain quantized orbits \ldots should play a special role, could well be granted; somewhat less easy to accept is the further assumption that the electrons moving on these curvilinear orbits \ldots radiate no energy. But that the sharply defined frequency of an emitted light quantum should be different from the frequency of the emitting electron would be regarded by a theoretician who had grown up in the classical school as {\it monstrous} and {\it almost inconceivable} \citep[pp.\ 150--151, our emphasis]{Lakatos}. 
\end{quotation}
Unfortunately, this passage is nowhere to be found in \citep{Born 1948}! 

One area of the old quantum theory in which the
``monstrous" element became glaringly and unavoidably apparent was in the treatment
of optical dispersion, the differential refraction of light of different colors. It was in this area that physicists most keenly felt the tension between orbital frequencies associated with individual states (the quantized electron orbits of the Bohr-Sommerfeld model) and  radiation frequencies associated with {\it transitions} between such states. One of the key points of Heisenberg's {\it Umdeutung} paper was to  formulate a new theory not in terms of properties of individual quantum states but  in terms of quantities associated with transitions between states {\it without even attempting to specify the states themselves}. What, above all, prepared the ground for this move, 
as we shall show in this section, was the development of a quantum theory of dispersion by Ladenburg, Reiche, Bohr, Kramers, and others. As Friedrich Hund (1896--1997) put it in his concise but rather cryptic history of quantum theory:
\begin{quotation}
In 1924 the question of the {\it dispersion of light} came to the foreground. It brought new points of view, and {\it it paved the way for quantum mechanics}
\citep[p.\ 128]{Hund 1984}.
\end{quotation}

By comparison, many of the other preoccupations
of the old quantum theory, such as a detailed understanding of spectral lines, the Zeeman and Stark effects, and the extension of the Bohr-Sommerfeld model to multi-electron atoms (in particular, helium)
mostly added to the overall confusion 
and did little to stimulate the shift to the new mode of thinking exemplified by the
{\it Umdeutung} paper.\footnote{For detailed analyses of some of these bewildering developments, see, e.g., \citep{Serwer 1977, Forman 1968, Forman 1970}.}   

The same is true---{\it pace} Roger \citet{Stuewer 1975}---for the broad acceptance of Einstein's 1905 light-quantum hypothesis following the discovery of the Compton effect in 1923.  What {\it was} crucial for the development of matrix mechanics were the $A$ and $B$ coefficients for emission and absorption of the quantum theory of radiation of \citep{Einstein 1916a, Einstein 1916b, Einstein 1917a}, {\it despite} its use of light quanta. Physicists working on dispersion theory were happy to use the $A$ and $B$ coefficients but they were just as happy to continue thinking of light as consisting of waves rather than particles. John \citet{Hendry 1981} makes the provocative claim that ``since Sommerfeld was the only known convert to the light-quantum concept as a result of the Compton effect whose opinions were of any real historical importance, this places Stuewer's thesis on the importance of the effect in some doubt" (p.\ 197). It is our impression that the Compton effect {\it did} convince many physicists of the reality of light quanta, just as Stuewer says it did,  but we agree with \citet[p.\ 6]{Hendry 1981} that this made surprisingly little difference for the further development of quantum physics.

\subsection{The classical dispersion theory of Lorentz and Drude}

Optical dispersion can boast of a venerable history in the annals of science reaching back at least to Descartes' rainbow and Newton's prism. The old quantum theory was certainly not the first theory for which dispersion presented serious difficulties. Neither Newtonian particle theories of light of the 18th century nor the wave theory of the early-19th century provided convincing accounts of the phenomenon \citep{Cantor 1983}. In the wave theory of Thomas Young (1773--1829) and Augustin Jean Fresnel (1788--1827), the index of refraction is related to the density of the luminiferous ether, the medium thought to carry light waves, inside transparent matter. Dispersion, the dependence of the index of refraction on frequency, thus forced proponents of the theory to assume that the ether density inside transparent matter was different for different colors! Similarly, one had to assume that matter contained different amounts of ether for the ordinary and the extraordinary ray in double refraction. The problem likewise affected the optics of moving bodies \citep{Janssen and Stachel 2004, Stachel 2005}. To account for the absence of any signs of motion of the earth with respect to the ether, \citet{Fresnel 1818}  introduced what is known as the ``drag" coefficient. He assumed that transparent matter with index of refraction $n$ carries along the ether inside of it with a fraction  $f = 1 - 1/n^2$ of its velocity with respect to the ether. Although it was widely recognized in the 19th century that the drag coefficient was needed to account for the null results of numerous ether drift experiments, many physicists expressed strong reservations about the underlying physical mechanism proposed by Fresnel, since it implied that, because of dispersion, matter had to drag along a different amount of ether for every frequency of light! 

One of the great triumphs of Lorentz's elaboration of the electromagnetic theory of light in the early 1890s was that he could derive the drag coefficient without having to assume an actual ether drag \citep{Lorentz 1892}. In Lorentz's theory, the ether is immobile, the ether density is the same everywhere, inside or outside of matter, and the index of refraction is related, not to ether density, but to the polarization of harmonically-bound charges, later to be identified with electrons, inside transparent matter. The resonance frequencies of these oscillating charges correspond to the material's experimentally-known absorption lines. Lorentz's dispersion theory was further developed by Paul Drude (1863--1906) (see, e.g., Drude, 1900, Pt.\ II, Sec.\ II, Ch.\ V). This classical Lorentz-Drude dispersion theory was remarkably successful in accounting for the experimental data. In 1896, Lorentz was able to account for the Zeeman effect on the basis of this same picture of the interaction of matter and radiation, which won him the 1902 Nobel prize \citep{Kox 1997}. Two centuries after Newton, there finally was a reasonably satisfactory theory of dispersion. Only two decades later, however, the model of matter underlying this theory was called into question again with the rise of the old quantum theory. The electrons oscillating inside atoms in the Lorentz-Drude model were replaced by electrons orbiting the nucleus in the Rutherford-Bohr model. As we shall see, the Lorentz-Drude theory nonetheless played a key role in the development of a quantum theory of dispersion in the early 1920s. 

The basic model of dispersion in the Lorentz-Drude theory is very simple.\footnote{Classical dispersion theory is covered elegantly in ch.\ 31 of Vol.\ 1 of the Feynman lectures (see also ch.\ 32 of Vol.\ 2). Feynman makes it clear that this theory remains relevant in modern physics: ``we will assume that the atoms are little oscillators, that is that the electrons are fastened elastically to the atoms \ldots You may think that this is a funny model of an atom if you have heard about electrons whirling around in orbits. But that is just an oversimplified picture. The correct picture of an atom, which is given by the theory of wave mechanics, says that, {\it so far as problems involving light are concerned}, the electrons behave as though they were held by springs" \citep[Vol.\ 1, sec.\ 31-4]{Feynman}. \label{feynman}}  Suppose an electromagnetic wave of frequency $\nu$ (we are not concerned with how and where this wave originated) strikes a charged one-dimensional simple harmonic oscillator with characteristic frequency $\nu_0$. We focus on the case of so-called {\it normal dispersion}, where the frequency $\nu$ of the electromagnetic wave is far from the resonance frequency $\nu_0$ of the oscillator. The case where $\nu$ is close to $\nu_0$ is called {\it anomalous dispersion}.  We can picture the oscillator as a point particle with mass $m$ and charge $-e$ (where $e$ is the {\it absolute value} of the electron charge) on a spring with equilibrium position $x=0$ and spring constant $k$, resulting in a restoring force $F=-kx$. The characteristic angular frequency $\omega_0 = 2 \pi \nu_0$ is then given by $\sqrt{k/m}$.
The electric field $E$ of the
incident electromagnetic wave\footnote{We need not worry about the effects of the magnetic field $B$. The velocity of electrons  in typical atoms is of order $\alpha c$, where $c$ is the velocity of light and $\alpha\simeq \frac{1}{137}$ is the fine-structure constant.  The effects due to the magnetic field are thus a factor $\frac{1}{137}$ smaller than
those due to the electric field and can be 
ignored in all situations considered in this paper.}   will induce an additional component of the motion at the imposed 
frequency $\nu$. This component will be superimposed on any preexisting oscillations at the characteristic frequency $\nu_0$
of the unperturbed system. It is this additional component of the particle motion, coherent 
 with the
incident wave (i.e., oscillating with frequency $\nu$), that is responsible for the secondary radiation that gives rise to dispersion. The time-dependence of this component is given by:
\begin{equation}
\Delta x_{\rm coh}(t) = A  \cos{ \omega t},
\label{coh}
\end{equation}
where $\omega = 2 \pi \nu$.
To determine the amplitude $A$, we substitute eq.\ (\ref{coh}) into the equation of motion for the system. As long as we are far from resonance, radiation damping can be ignored and the equation of motion is simply:\footnote{In sec.\ 5.3,
we show how to take into account the effects of radiation damping.} 
\begin{equation}
\label{eq:2.14}
m\ddot{x} = -m \omega_0^2  x - e E \cos{ \omega t},
\end{equation}
where dots indicate time derivatives and where we have made the innocuous simplifying assumption that the electric field of the incident wave is in the $x$-direction. Substituting $\Delta x_{\rm coh}(t)$ in eq.\ (\ref{coh}) for $x(t)$ in eq.\ (\ref{eq:2.14}), we find:
\begin{equation}
\label{eq:2.15}
-m \omega^2 A \cos{\omega t} = (- m \omega_0^2 A - e E) \cos{\omega t}.
\end{equation}
It follows that
\begin{equation}
\label{eq:2.16}
A=\frac{e E}{m(\omega^2 - \omega_0^2)}.
\end{equation}
The central quantity in the Lorentz-Drude dispersion theory is
the dipole moment $p(t) \equiv - e \Delta x_{\rm coh}(t)$ of the oscillator induced by the electric field of the incident electromagnetic wave. From eqs.\ (\ref{coh}) and (\ref{eq:2.16}) it follows that:
\begin{equation}
\label{eq:1.1}
p(t) =  -e \Delta x_{\rm coh}(t) = \frac{e^2 E}{4\pi^2m(\nu_0^2 - \nu^2)} \cos{2 \pi \nu t}.
\end{equation}
For groups of $n_{i}$ oscillators of characteristic frequencies $\nu_i$ per unit volume, this formula for the dipole moment naturally generalizes to
the following result for the polarization (i.e., the dipole moment per unit volume):
\begin{equation}
\label{eq:1.2}
P(t)  = \frac{e^2 E}{4\pi^2m} \sum_i \frac{n_i}{\nu_i^2 - \nu^2} \; \cos{2 \pi \nu t}.
\end{equation}
The number of oscillators of characteristic frequency $\nu_i$ will be some fraction $f_i$ of the numbers of atoms in the volume under consideration. This fraction was often called the `oscillator strength' in the literature of the time.
The polarization $P$ determines the index of refraction $n$
(see, e.g., Feynman {\it et al.}, 1964, Vol.\ 1, 31-5). The agreement of eq.\,(\ref{eq:1.2}) with the data from experiments on dispersion was not perfect, but dispersion was nonetheless seen as an important success for Lorentz's classical theory. 

\subsection{The Sommerfeld-Debye theory and its critics}

An early and influential attempt to bring dispersion theory under the umbrella of the old quantum theory was made by  \citet{Sommerfeld 1915b, Sommerfeld 1917} and by his former student Peter Debye (1884--1966) \citep{Debye 1915}.\footnote{\label{dispersion-lit} For other historical discussions of the development of quantum dispersion theory, see, e.g., \citep[pp.\ 224--230]{Darrigol}, \citep[pp.\ 146--159, pp.\ 215--222]{Dresden}, \citep[p.\ 165 and sec.\ 4.3, especially pp.\ 188--195]{Jammer 1966}, \citep[Vol.\ 1, sec.\ VI.1; Vol.\ 2, sec.\ III.5, pp.\ 170--190; Vol.\ 6, sec.\ III.1 (b), pp.\ 348--353]{Mehra Rechenberg}, and \citep[Vol.\ 1, p.\ 401; Vol.\ 2, pp.\ 200--206]{Whittaker}. \citet[sec.\ 49, pp.\ 156--159]{Van Vleck 1926a} briefly discusses the early attempts to formulate a quantum theory of dispersion in his review article on the old quantum theory. We focus on the theory of Debye and Sommerfeld. Van Vleck also mentions theories by Charles Galton Darwin (1887--1962), Adolf Gustav Smekal (1895--1959), and Karl F.\ Herzfeld (1892--1978). All three of these theories make use of light quanta. In addition, strict energy conservation is given up in the theory of \citet{Darwin 1922, Darwin 1923}, while in the theories of \citet{Smekal 1923} and \citet{Herzfeld 1924} orbits other than those picked out by the Bohr-Sommerfeld condition are allowed, a feature known as  ``diffuse quantization." For other (near) contemporary reviews of dispersion theory, see \citep[pp.\ 86--96]{Pauli 1926}, \citep[pp. 669--682]{Andrade 1927}, and \citep{Breit 1932}.  \citet[pp.\ 17--18]{Stolzenburg 1984} briefly discusses Bohr's critical reaction to Darwin's dispersion theory.} Clinton J.\ Davisson (1881--1958), then working at the Carnegie Institute of Technology in Pittsburgh, also contributed \citep{Davisson 1916}.\footnote{In 1927 at Bell Labs, Davisson and his assistant Lester H.\ Germer (1896--1971) would do their celebrated work on electron diffraction \citep{Davisson and Germer 1927}, another great American contribution to (experimental) quantum physics for which the authors received the 1937 Nobel prize \citep[pp.\ 188--189]{Kevles 1978}.}  The Debye-Sommerfeld theory, as it came to be known, was based on the dubious assumption that the secondary radiation coming from small perturbations of a Bohr orbit induced by incident radiation could be calculated on the basis of ordinary  classical
electrodynamics, even though, by the basic tenets of the Bohr model, the classical theory did {\it not} apply to the original unperturbed orbit. In other words, it was assumed that, while the large accelerations of electrons moving on Bohr orbits  would  produce no radiation whatsoever, the comparatively small accelerations involved in the slight deviations from these orbits caused by weak incident radiation would produce radiation.\footnote{\citet[p.\ 502]{Sommerfeld 1915b} realized that this assumption was problematic and tried (unconvincingly) to justify it.} Otherwise, the theory stayed close to the Lorentz-Drude theory, substituting small deviations in the motion of electrons from their Bohr orbits for small deviations from the vibrations of simple harmonic oscillators at their characteristic frequencies.

Both the Swedish physicist Carl Wilhelm Oseen (1879--1944) and Bohr severely criticized the way in which Debye and Sommerfeld modeled their quantum dispersion theory on the classical theory. \citet{Oseen 1915} wrote: ``Bohr's atom model can in no way be reconciled with the fundamental assumptions of Lorentz's electron theory. We have to make our choice between these two theories" (p.\ 405).\footnote{Quoted and discussed in \citep[Vol.\ 2, p.\ 337]{Bohr 1972--1996}}
Bohr agreed. The central problem was that in Bohr's theory the link between radiation frequencies and orbital frequencies had been severed. As Bohr explained to Oseen in a letter of December 20, 1915, if the characteristic frequencies involved in dispersion
\begin{quotation}
\ldots are determined by the laws for quantum emission, the dispersion cannot, whatever its explanation, be calculated from the motion of the electrons and the usual electrodynamics, which does not have the slightest connection with the frequencies considered \citep[Vol.\ 2, p.\ 337]{Bohr 1972--1996}.
\end{quotation}
Bohr elaborated on his criticism of the Debye-Sommerfeld theory in a lengthy paper intended for publication in {\it Philosophical Magazine} in 1916 but withdrawn after it was already typeset.\footnote{It can be found in \citep[Vol.\ 2, pp.\ 433--461]{Bohr 1972--1996}. For further discussion of Bohr's early views on dispersion, see \citep[pp.\ 281--283]{Heilbron and Kuhn}.} Bohr argued (we leave out the specifics of the experiments on dispersion in various gases that Bohr mentions in this passage): 
\begin{quotation}
[E]xperiments \ldots show that the dispersion \ldots can be represented with a high degree of approximation by a simple Sellmeier formula\footnote{This is a formula of the form of eq.\ (\ref{eq:1.2}) derived on the basis of an elastic-solid theory of the ether in \citep{Sellmeier 1872} \citep[p.\ 189]{Jammer 1966}.} in which the characteristic frequencies coincide with the frequencies of the lines in the \ldots spectra \ldots 
[T]hese frequencies correspond with transitions between the normal states of the atom \ldots On this view we must consequently assume that the dispersion \ldots depends on the same mechanism as the transition between different stationary states, and that it cannot be calculated by application of ordinary electrodynamics from the configuration and motions of the electrons in these states \citep[Vol.\ 2, pp.\ 448--449]{Bohr 1972--1996}.
\end{quotation}
In the next paragraph, Bohr added a prescient comment. Inverting the line of reasoning in the pasage above that dispersion should depend on the same mechanism as  transitions between states, he suggested that transitions between states, about which the Bohr theory famously says nothing, should depend on the same mechanism as dispersion:  ``[i]f the above view is correct \ldots we must, on the other hand, assume that this mechanism [of transitions between states] shows a close analogy to an ordinary electrodynamic vibrator" (ibid.).

As we shall see, in the quantum dispersion theory of the 1920s,  the Lorentz-Drude oscillators were grafted onto the Bohr model. For the time being, however, it was unclear how to arrive at a satisfactory quantum theory of dispersion. The quasi-classical Debye-Sommerfeld theory led to a formula for the induced polarization of the form of eq.\ (\ref{eq:1.2}) but with resonance poles at the orbital frequencies.
As Oseen and Bohr pointed out, this was in blatant contradiction with the experimental data, which clearly indicated that the poles should be at the radiation frequencies, which in Bohr's theory differed sharply from the orbital frequencies.

This criticism is repeated in more sophisticated form in a paper by Paul Sophus Epstein (1883--1966) with the subtitle ``Critical comments on dispersion." This paper is the concluding installment of a trilogy on the application of classical perturbation theory to problems in the old quantum theory \citep{Epstein 1922a, Epstein 1922b, Epstein 1922c}. Epstein, a Russian Jew who studied with Sommerfeld in Munich, was the first European quantum theorist to be lured to America. In 1921 Millikan brought him to the California Institute of Technology in Pasadena, despite prevailing antisemitic attitudes \citep[pp.\ 211--212]{Kevles 1978}.\footnote{For further discussion of Epstein's position at Caltech, see  \citep[pp.\ 507--520]{Seidel 1978}.}
In his 1926 review article Van Vleck emphasizes the importance of the work of his colleague at Caltech and notes that it ``is rather too often overlooked" \citep[p.\ 164, note 268]{Van Vleck 1926a}, to which one might add: ``by European physicists." As we saw in sec.\ 2.4, Van Vleck felt the same way about his own contributions. Like Van Vleck, Epstein apparently complained about this lack of recognition to Born. This can be inferred from a letter from Born to Sommerfeld of January 5, 1923, shortly before a visit of the latter to the United States:
\begin{quotation}
When you talk to Epstein in Pasadena and he complains about me, tell him that he should show you the very unfriendly letter he wrote to me because he felt that his right as first-born had been compromised by the paper on perturbation theory by Pauli and me [Born and Pauli, 1922, which appeared shortly after Epstein's trilogy]. Also tell him that I do not answer such letters but that I do not hold a grudge against him because of his impoliteness (to put it mildly) \ldots In terms of perturbative quantization we are ahead of him anyway \citep[p.\ 137]{Sommerfeld 2004}.\footnote{This letter is quoted and discussed in \citep[p.\ 96]{Eckert 1993}} 
\end{quotation}
To deal with the kind of multiply-periodic systems that represent hydrogenic atoms (i.e., atoms with only one valence electron) in the old quantum theory, Epstein customized techniques developed in celestial mechanics for computing the
 perturbations of the orbits of the inner planets due to the gravitational pull of the outer ones.\footnote{One of the sources cited by \citet[p.\ 216]{Epstein 1922a} is \citep{Charlier}. This source is also cited in \citep[p.\ 114]{Bohr 1918}, \citep[p.\ 8]{Kramers 1919}, and \citep[p.\ 154]{Born and Pauli 1922}. In their interviews for the AHQP, both Van Vleck (p.\ 14 of the transcript of session 1) and Heisenberg (p.\ 24 of the transcript of session 5) mention that they studied Charlier as well.} The perihelion advance of Mercury due to such perturbations, for instance, is more than ten times the well-known $43^{\prime \prime}$ per century due to the gravitational field of the sun as given by  general relativity. Such calculations in classical mechanics are also the starting point of the later more successful approach to dispersion theory by Kramers and Van Vleck. Epstein clearly recognized that these calculations by themselves do not lead to a satisfactory theory of dispersion. 
In the introduction of his paper, \citet[p.\ 92]{Epstein 1922c} explains that he discusses dispersion mainly because it nicely illustrates some of the techniques developed in the first two parts of his trilogy. He warns the reader  that he is essentially following the Debye-Sommerfeld theory, and emphasizes that ``this point of view leads to internal contradictions so strong that I consider the Debye-Davysson [sic] dispersion theory [as Epstein in Pasadena referred to it] to be untenable" (ibid.). The central problem is once again the discrepancy between radiation frequencies and orbital frequencies. As Epstein wrote in the conclusion of his paper:
\begin{quotation}
the positions of maximal dispersion and absorption [in the formula he derived] do not lie at the position of the emission lines of hydrogen but at the position of the mechanical frequencies of the model \ldots {\it the conclusion seems  unavoidable to us  that the foundations of the Debye-Davysson} [sic] {\it theory are incorrect} \citep[pp.\ 107--108]{Epstein 1922c}.
\end{quotation}
Epstein recognized that a fundamentally new approach was required: ``We believe that \ldots dispersion theory must be put on a whole new basis, in which one takes the Bohr frequency condition into account from the very beginning" (ibid., p.\ 110). 

\subsection{Dispersion in Breslau: Ladenburg and Reiche}

Unbeknownst to Epstein, quantum dispersion theory had already begun to emerge from the impasse he called attention to in 1922. The year before, Ladenburg had introduced one of two key ingredients needed for a satisfactory treatment of dispersion in the old quantum theory:  the emission and absorption coefficients of Einstein's quantum theory of radiation. The other critical ingredient, as we shall see below, was Bohr's correspondence principle. 

Ladenburg spent most of his career doing experiments on dispersion in gases, beginning in 1908, about two years after he joined the physics department, then headed by Otto Lummer (1860--1925), at the University of Breslau, his hometown \citep{Ladenburg 1908}.\footnote{See the entry on Ladenburg by A.\ G.\ \citet{Shenstone 1973} in the {\it Dictionary of Scientific Biography}.} He stayed in Breslau until 1924, when he accepted a position at the {\it Kaiser Wilhelm Institut} in Berlin. There he continued his work with the help of students such as Hans Kopfermann (1895--1963), Agathe Carst, S.\ Levy, and G.\ Wolfsohn. Ladenburg and his group reported the results of their experiments on dispersion in a series of papers published between 1926 and 1934.\footnote{See \citep[Vol.\ 6, Ch.\ 3(b), pp.\ 348--353]{Mehra Rechenberg} and \citep[p.\ 555]{Shenstone 1973} for detailed references and brief discussions.} Ladenburg's direct involvement ceased with his emigration  to the United States in 1931. 

Ladenburg and Stanislaw Loria (1883--1958) had established  early on that the frequency of the $H_\alpha$ line in the Balmer series in the hydrogen spectrum corresponds to a pole in the Lorentz-Drude dispersion formula \citep[p.\ 866]{Ladenburg and Loria 1908}. Given that the Sommerfeld-Debye theory flies in the face of this experimental fact, Ladenburg was never attracted to that theory. He simply kept using a dispersion formula with poles at the observed radiation frequencies. He focused on the numerator rather than the denominator of the dispersion formula. This is made particularly clear in the AHQP interviews with two of his collaborators in the early 1920s---Rudolph Minkowski (1895--1976), a nephew of Hermann Minkowski, who took his doctorate under Ladenburg in 1921 and co-authored \citep{Ladenburg and Minkowski 1921};
and Fritz Reiche, who came to Breslau to replace Schr\"odinger in 
1921.\footnote{The following information is based on an autobiographical statement by Reiche published as an appendix to \citep{Bederson 2005}.} After his doctorate (with Planck) in Berlin  in 1907, Reiche had already spent three years in Breslau and had become close friends  with Ladenburg during that period. He had gone back to Berlin in 1911. When he returned to Breslau ten years later, he stayed until he was dismissed in 1933.\footnote{It was not until 1941 that he finally managed to emigrate to the United States.} Reiche's help is prominently acknowledged in \citep[p.\ 140, note]{Ladenburg 1921}. Ladenburg was first and foremost an experimentalist and he welcomed input from his theoretician friend and colleague.\footnote{Asked by Kuhn whether Ladenburg was ``strictly an experimentalist," Rei\-che said: ``He was, as far as I understand, a very good experimental man, but he was one of the men who could make, let me say, easy theoretical work" (p.\ 10 of the transcript of the last of three sessions of the interview).}  The two of them co-authored a pair of follow-up papers \citep{Ladenburg and Reiche 1923, Ladenburg and Reiche 1924}. Discussing the first of these, Reiche told Kuhn and Uhlenbeck in 1962:
\begin{quotation}
we did not derive a consistent dispersion theory, in which instead of the revolution numbers the emitted lines came out. We thought it completely self-evident, that one had to change the denominator of the dispersion formula in such a way that the frequencies were the emitted line frequencies, and not something which has to do with (the orbit) [sic].\footnote{P.\ 11 of the transcript of the second of three sessions of the AHQP interview with Reiche.}
\end{quotation}  
``But that was a big step," Uhlenbeck interjected, ``wasn't it?" ``But not in this direction," Reiche insisted, 
\begin{quotation}
Only in the direction of explaining that the $N$ which is on top of the dispersion formula---the number of [dispersion] electrons [sic]. It never came out correctly equal to the number of atoms, or to the number of atoms multiplied by the number of electrons in an atom. It gave, under certain conditions, even numbers which are less than the whole number of atoms. They were written very often with a German $N$ \ldots This was the main aim of the whole thing [Ladenburg and Reiche, 1923]. There, based on a previous paper by Ladenburg [1921], we found a relation between the German $N$ and the real number of atoms. The $f$ were not 1 or 2 or 3 or something like this, but could be point 5 or the like. And the explanation of this was the aim of this dispersion paper. But it did not come out that we had a correct and consistent theory in which the denominator gave now the emitted frequencies. This, I think, was only done by Kramers [1924a, b], first of all.\footnote{Ibid.\ (the first set of square brackets are in the transcript). Dispersion is discussed at greater length during the third session of the interview (see pp.\ 10--14 of the transcript).}
\end{quotation}
Ladenburg's dispersion experiments had indicated all along that the oscillator strength $f_i$,
the number of dispersion electrons with characteristic frequency $\nu_i$ per number of atoms, was not on the order of unity, as one would expect on the basis of the Lorentz-Drude theory, but much smaller. For the frequency $\nu_i$ corresponding to the $H_\alpha$ line in the Balmer series in the hydrogen spectrum, for instance, \citet[p.\ 865]{Ladenburg and Loria 1908} found that there was only 1 dispersion electron per 50,000 molecules, and they cited findings of 1 dispersion electron per 200 molecules in sodium vapor. Such low values were quite inexplicable on classical grounds. In the Bohr model the $H_\alpha$ (absorption)  line corresponds to a transition from the $n=2$ to the $n=3$ state of the hydrogen atom. That Ladenburg found such a low value for what he interpreted classically as the number of dispersion electrons at the frequency of the $H_\alpha$ line is explained in Bohr's theory simply by noting that only a tiny fraction of the atoms will be in the $n=2$ state \citep[p.\ 156]{Ladenburg 1921}. Ladenburg's key contribution was that he recognized that the oscillator strengths corresponding to various transitions could all be interpreted in terms of transition probabilities, given by Einstein's $A$ and $B$ coefficients. Hence the title of his paper: ``The quantum-theoretical interpretation of the number of dispersion electrons" \citep{Ladenburg 1921}. 


Ladenburg obtained a relation between the oscillator strengths and the $A$ and $B$ coefficients  by equating results derived for what would seem to be two mutually exclusive models of matter, a classical and a quantum model. He calculated the rate of absorption of energy both for a collection of classical oscillators \`{a} la Lorentz and Drude, resonating at the absorption frequencies, and for a collection of atoms \`{a} la Bohr and Einstein with transitions between discrete energy levels corresponding to these same frequencies. Ladenburg set the two absorption rates equal to one another. His paper only gives the resulting expression for the numerator of the dispersion formula. Combining Ladenburg's theoretical relation between classical oscillator strengths and quantum transition probablities with his experimental evidence that the resonance poles should be at the radiation frequencies, we arrive at the following formula---in our notation, based on \citep{Van Vleck 1924b}---for the induced polarization of a group of $N_{r}$ atomic systems in
their ground state $r$
 \begin{equation}
 \label{eq:1.3}
     P_{r}(t) = \frac{N_{r}c^{3}E}{32\pi^{4}}\sum_{s}\frac{A_{s\rightarrow r}}{\nu_{s\rightarrow r}^{2}(\nu_{s\rightarrow r}^{2}-\nu^{2})}\cos{2\pi\nu t},
 \end{equation}
where $\nu_{s\rightarrow r}$
is the frequency for a transition from the excited states $s$ to $r$ and $A_{s\rightarrow r}$  is Einstein's emission coefficient  for this transition.

Ladenburg's paper initially did not attract much attention. It is not mentioned in Epstein's trilogy the following year, but then Epstein was working in faraway California. More surprisingly, quantum physicists in G\"ottingen, Munich, and Copenhagen, it seems, also failed to take notice, even though Ladenburg was well-known to his G\"ottingen colleagues Born and James Franck (1882--1964). Ladenburg had actually prevented that Born, a fellow Breslau native, was sent to the trenches in World War I. Ladenburg had recruited Born for an army unit under his command in Berlin devoted to artillery research \citep[pp.\ 71--72]{Greenspan}. Bohr and Ladenburg also knew  each other personally: Ladenburg had attended Bohr's colloquium in Berlin in April 1920 and the two men had exchanged a few letters since \citep[Vol.\ 4, pp.\ 709--717]{Bohr 1972--1996}.

Heisenberg later attributed the neglect of Ladenburg in G\"{o}ttingen and Munich to the problem of connecting Ladenburg's work, closely tied to Einstein's radiation theory, to the dominant Bohr-Sommerfeld theory.\footnote{See p.\ 8 of the transcript of session 4 of the AHQP interview with Heisenberg, parts of which can be found in \citep[Vol.\ 2, pp.\ 175--176]{Mehra Rechenberg}, although the authors cite their own conversations with Heisenberg as their source.\label{plagiarism2}} 
According to Heisenberg, it was only when \citet{Kramers 1924a, Kramers 1924b} re\-derived Ladenburg's formula as a special case of his own more general dispersion formula that  its significance was widely appreciated.\footnote{Jordan had the same impression (see pp.\ 24--25 of the transcript of the first session of Kuhn's interview with Jordan for the AHQP in June 1963). It also fits with Born's recollections. In his autobiography, \citet{Born 1978} notes:  ``An important step was made by my old friend from Breslau \ldots Ladenburg \ldots A detailed account was given by Ladenburg and Reiche, my other old friend from Breslau \ldots On the basis of these investigations, Kramers \ldots succeeded in developing a complete `dispersion formula'" (pp.\ 215--216).} 
Ladenburg's own derivation had been unconvincing, at least to most physicists.\footnote{As Kuhn put it in his AHQP interview with Slater: ``Of course, there was a good deal that appeared to most physicists as pretty totally {\it ad hoc} about the Reiche-Ladenburg work, and the whole question as to why it was the transition frequencies that occurred in the denominator rather than the orbital frequencies." Slater disagreed: ``This seemed to me perfectly obvious \ldots" (p.\ 41 of the transcript of the first session of the interview).}  In addition to just assuming the poles in the dispersion formula to be at the radiation frequencies rather than at the orbital frequencies, Ladenburg offered no justification for equating the rate of energy absorption in his classical model to that for his Einsteinian quantum model of matter. \citet[p.\ 10]{Van der Waerden} suggests that Ladenburg appealed to Bohr's correspondence principle in his derivation of the relation between oscillator strengths and $A$ and $B$ coefficients, but the correspondence principle is not mentioned anywhere in Ladenburg's paper. The full dispersion formula (\ref{eq:1.3})---admittedly only implicit in Ladenburg's paper but associated with it, not just by later historians but also by his contemporaries---can certainly not be derived with the help of the correspondence principle, since it only holds for atoms in their ground state and {\it not} for atoms in highly excited states where classical and quantum theory
may be expected to merge in the sense of the correspondence principle. Still, if Heisenberg's later recollections are to be trusted, it might have helped the reception of Ladenburg's paper had he made some reference to the correspondence principle.

Unlike his colleagues in G\"ottingen and Munich, Bohr in fact took notice of Ladenburg's paper early on. He was just slow, as usual, to express himself about it in print. As noted in \citep[p.\ 192]{Hendry 1981}, Bohr referred to \citep{Ladenburg 1921} in the very last sentence of a manuscript he did not date but probably started and abandoned in 1921 \citep[Vol.\ 3, pp.\ 397--414]{Bohr 1972--1996}.  In a paper submitted in November 1922, \citet[p.\ 162]{Bohr 1923b} finally discussed Ladenburg's work in print. After repeating some of the observations about dispersion made in the passages of his unpublished 1916 paper quoted in sec.\ 3.2, Bohr, in his tortuous verbose style, made some highly interesting remarks that anticipate aspects of the BKS theory of 1924 (see sec.\ 4):
\begin{quotation}
the phenomena of dispersion must thus be so conceived that the reaction of the atom on being subjected to radiation is closely connected with the unknown mechanism which is answerable [the German has {\it verantwortlich}: responsible]  for the emission of the radiation on the transition between stationary states. In order to take account of the observations, it must be assumed that this mechanism \ldots becomes active when the atom is illuminated in such a way that the total reaction of a number of atoms is the same as that of a number of harmonic oscillators in the classical theory,\footnote{Note the similarity between Bohr's description here to Feynman's observation (quoted in note \ref{feynman} above) that atoms behave like oscillators ``so far as problems involving light are concerned."} the frequencies of which are equal to those of the radiation emitted by the atom in the possible processes of transition, and the relative number of which is determined by the probability of occurrence of such processes of transition under the influence of illumination. A train of thought of this kind was first followed out closely in a work by Ladenburg [1921] in which he has tried, in a very interesting and promising manner, to set up a direct connection between the quantities which are important for a quantitative description of the phenomena of dispersion according to the classical theory and the coefficients of probability appearing in the deduction of the law of temperature radiation by Einstein
\citep[Vol.\ 3, p.\ 496]{Bohr 1972--1996}.
\end{quotation}  
A letter from Bohr to Ladenburg of May 17, 1923 offers further insights into Bohr's developing views on the mechanism of radiation:
\begin{quotation}
to interpret the actual observations, it \ldots seems necessary to me that the quantum jumps are not the direct cause of the absorption of radiation, but that they represent an effect  which accompanies the continuously dispersing (and absorbing) effect of the atom on the radiation, even though we cannot account in detail for the quantitative relation [between these two effects] with the usual concepts of physics \citep[Vol.\ 5, p.\ 400]{Bohr 1972--1996}.
\end{quotation}
At the beginning of this letter, Bohr mentioned the vagueness of some of his earlier pronouncements on the topic. After the passage just quoted he acknowledged ``that these comments are not far behind the earlier ones in terms of vagueness. I do of course reckon with the possibility that I am on the wrong track but, if my view contains even a kernel of truth, then it lies in the nature of the matter that the demand for clarity in the current state of the theory is not easily met" (ibid.). Bohr need not have been so apologetic. His comments proved to be an inspiration to Ladenburg and Reiche. On June 14, 1923, Ladenburg wrote to Bohr:
\begin{quotation}
Over the last few months Reiche and I have often discussed [the absorption and scattering of radiation] following up on your comments in [Bohr 1923b] about reflection and dispersion phenomena and on my own considerations [Ladenburg 1921] which you were kind enough to mention there  \citep[Vol.\ 5, pp.\ 400--401]{Bohr 1972--1996}.
\end{quotation}
In this same letter, Ladenburg announced his forthcoming paper with Reiche in a special issue of {\it Die Naturwissenschaften} to mark the tenth anniversary of Bohr's atomic theory. In the conclusion of this paper, they wrote:\footnote{Quoted and discussed in \citep[p.\ 192]{Hendry 1981}.}
\begin{quotation}
Surveying the whole area of scattering and dispersion discussed here, we have to admit that we do not know the real [{\it eigentlich}] mechanism through which an incident wave acts on the atoms and that we cannot describe the reaction of the atom in detail. This is no different by the way in the case of the real [{\it eigentlich}] quantum process, be it that an external wave $\nu_0$ lifts electrons into higher quantum states, or be it that a wave $\nu_0$ is sent out upon the return to lower states. We nevertheless believe on the grounds of the observed phenomena that the end result of a process in which a wave of frequency $\nu$ acts upon the atom should not be seen as fundamentally different from the effect that such a wave exerts on classical oscillators \citep[p.\ 597]{Ladenburg and Reiche 1923}.
\end{quotation}
\citet[p.\ 588, p.\ 590]{Ladenburg and Reiche 1923} introduced the term ``substitute oscillators" [{\it Ersatzoszillatoren}] for such classical oscillators representing the atom as far as its interaction with radiation is concerned. They credited Bohr with the basic idea.\footnote{See also \citep[p.\ 672]{Ladenburg and Reiche 1924}. \citet[p.\ 159, note 260]{Van Vleck 1926a} reports that Lorentz made a similar suggestion at the third Solvay congress in 1921 \citep[p.\ 24]{Verschaffelt 1923}, but does not mention Ladenburg and Reiche in this context, attributing the idea to \citep{Slater 1924} instead.} As we shall see in sec.\ 4, these substitute oscillators became the virtual oscillators of BKS. \citet[p.\ 672]{Ladenburg and Reiche 1924} themselves noted the following year that substitute oscillators were now called virtual oscillators \citep[p.\ 141]{Konno 1993}. The Berlin physicist Richard Becker (1887--1955) likewise noted in a paper written in the context of BKS the following year: ``these virtual oscillators are substantially identical with the `substitute oscillators' already introduced by Ladenburg and Reiche" \citep[p.\ 174, note 2]{Becker 1924}.\footnote{Quoted in \citep[p.\ 141]{Konno 1993}.} That same year, \citet[p.\ 350]{Herzfeld 1924} still used the term `substitute oscillators,' citing \citep{Ladenburg and Reiche 1923}. The term can also be found, without attribution, in the famous paper by \citet[p.\ 884]{Born and Jordan 1925b} on matrix mechanics.\footnote{We are grateful to J\"urgen Ehlers for drawing our attention to this passage, which is not in the part of \citep{Born and Jordan 1925b} included in \citep{Van der Waerden}.}

Unlike Ladenburg in 1921, Ladenburg and Reiche prominently mentioned both Bohr's atomic theory and the correspondence principle in their 1923 paper. The authors' understanding and use of the correspondence principle, however, are still tied strongly to Einstein's quantum theory of radiation. Their ``correspondence" arguments apply not to individual quantum systems, for which Bohr's correspondence principle was formulated, but to collections of such systems in thermal equilibrium with the ambient radiation.\footnote{That Ladenburg and Reiche did not carefully distinguish between individual systems and collections of such systems becomes more understandable if we bear in mind that they were trying to combine Einstein's quantum theory of radiation and Bohr's correspondence principle. These two elements  belong to two different strands in the development of quantum physics, characterized as follows in a concise and perceptive overview of the early history of quantum physics: ``The first approach, dominated by the Berlin physicists Einstein, Planck, and Nernst, and by \ldots Ehrenfest \ldots involved the thermodynamics properties of matter and the nature of radiation \ldots The other trend, centered socially in Copenhagen, Munich and G\"ottingen, consisted of the application of the quantum to individual atoms and molecules" \citep[p.\ 336]{Darrigol 2002}.} The authors also do not limit their ``correspondence" arguments to the regime of high quantum numbers \citep[especially secs.\ 4--5, pp.\ 586--589]{Ladenburg and Reiche 1923}. These problems invalidate many of the results purportedly derived from the correspondence principle in their paper. Drawing on earlier work by Planck, they derived a result for emission consistent with the correspondence principle (i.e., merging with the classical result in the limit of high quantum numbers), but their attempts to derive similar results for absorption and dispersion were unconvincing. In fact, it may well be that these dubious attempts inspired Van Vleck to formulate correspondence principles for emission and absorption himself (see sec.\ 5.3 for further discussion).

\subsection{The Kramers dispersion formula}

Given Bohr's strong interest in the subject, it is not surprising that  his first lieutenant Kramers took the next  big step in quantum dispersion theory.\footnote{\citet[p.\ 46]{Hendry 1984} goes as far as calling Kramers' theory ``the Bohr-Kramers dispersion theory."}
 Formula (\ref{eq:1.3}) based on Ladenburg's insights only holds for systems in the ground state. The correspondence principle only applies to highly excited states. \citet{Kramers 1924a, Kramers 1924b} found that the correspondence principle requires a formula with {\it two} terms.\footnote{In addition to the literature cited in note \ref{dispersion-lit}, see  \citep[pp.\ 23--30]{Ter Haar 1998} and, especially,  \citep{Konno 1993} for discussion of Kramers' work on dispersion theory.} In our notation---which once again follows \citep[p.\ 344, eq.\ 17]{Van Vleck 1924b}---the induced polarization $P_{r}$ of $N_r$ atoms in a state labeled by the quantum number $r$ is given by 
 \begin{equation}
 \label{eq:1.4}
     P_{r}(t) = \frac{N_{r}c^{3}E}{32\pi^{4}}\left(\sum_{s>r}\frac{A_{s\rightarrow r}}{\nu_{s\rightarrow r}^{2}(\nu_{s\rightarrow r}^{2}-\nu^{2})}-\sum_{t<r}\frac{A_{r\rightarrow t}}{\nu_{r\rightarrow t}^{2}(\nu_{r\rightarrow t}^{2}-\nu^{2})}\right)\cos{2\pi\nu t},
 \end{equation}
where $s$ and $t$ are the quantum numbers labeling states above and below $r$, respectively (see secs.\ 5.1--5.2 and 6.2 for detailed derivations). For high values of $r$ this formula merges with the classical result. In the spirit of the correspondence principle, Kramers took the leap of faith that it holds all the way down to low quantum numbers. If $r$ is the ground state, the second term vanishes and the Kramers formula (\ref{eq:1.4}) reduces to the Ladenburg formula (\ref{eq:1.3}). Like \citet{Ladenburg and Reiche 1923}, Kramers interpreted his formula in terms of oscillators, distinguishing between ``absorption oscillators" for the first term and ``emission oscillators" for the second term \citep[pp.\ 179--180]{Kramers 1924a}. Kramers introduced the characteristic times $\tau_{i \rightarrow f}$ inversely proportional to $(e^2/m)\nu^2_{i \rightarrow f}$. So instead of  factors $\nu^2_{i \rightarrow f}$  in the denominators in the two terms in eq.\ (\ref{eq:1.4}), the formula given by \citet[p.\ 179, eq.\ 5]{Kramers 1924a} has factors $(e^2/m) \tau_{i \rightarrow f}$ in the numerators.\footnote{The polarization given by Kramers' formula is three times the polarization given by Van Vleck (i.e., by our  eq (\ref{eq:1.4})). This is because Kramers assumed that the vibrations in the atom are lined up with the electric field, whereas Van Vleck assumed the relative orientation of vibrations and fields to be random \citep[p.\ 344, note 25]{Van Vleck 1924b}.} Because of the minus sign in front of the second term, the emission oscillators appear to have negative mass, which is why Kramers also called them ``negative oscillators" (ibid.). \citet[p.\ 30, note 2]{Van Vleck 1924a} gave a more satisfactory interpretation of this minus sign, interpreting Kramers' formula for dispersion the same way he interpreted the formula he himself had found for absorption, as giving the net dispersion in a given quantum state as the difference between contributions from transitions to higher and transitions to lower states.  

Kramers initially only published two notes in {\it Nature} on his new dispersion formula  \citep{Kramers 1924a, Kramers 1924b}. Since these were submitted {\it after} \citep{BKS}, he used the new BKS terminology of `virtual oscillators' in both of them. As we shall see in sec.\ 4, this caused considerable confusion, both at the time and in the historical literature, about the relation between  BKS and dispersion theory. Kramers' notes, moreover, are short on detail. The first, submitted on March 25, contains only the briefest of hints as to how the new dispersion formula had been found. The second, submitted on July 22 in response to a letter by Minnesota's Gregory \citet{Breit 1924b}, contains at least an outline of the derivation. Kramers did not get around to publishing  the derivation in full until his paper with Heisenberg, completed over the Christmas break of 1924, received by {\it Zeitschrift f\"{u}r Physik} on January 5, 1925, and published two months later  \citep[Vol.\ 2, p.\ 181]{Mehra Rechenberg}. According to Slater, however, the basic results had been in place by the time he, Slater, arrived in Copenhagen in December 1923. After dissing Bohr in the letter to Van Vleck quoted in sec.\ 2.2, Slater goes on to say that
\begin{quotation}
Kramers hasn't got much done, either. You perhaps noticed his letter to
Nature on dispersion [Kramers, 1924a]; the formulas \& that he had before I came, although he
didn't see the exact application;  and except for that he hasn't done
anything, so far as I know. They seem to have too much administrative work
to do. Even at that, I don't see what they do all the time. Bohr hasn't
been teaching at all, Kramers has been giving one or two courses.\footnote{Slater to Van Vleck, July 27, 1924 (AHQP).}
\end{quotation}
Part of what kept Kramers from his work in early 1924, as can be gathered from correspondence with Ladenburg and Reiche, was that his wife had fallen ill. In 1923, the Breslau physicists had already exchanged a few letters about dispersion with their colleague in Copenhagen.\footnote{See Reiche to Kramers, May 9, 1923 and December 28, 1923, and Ladenburg to Kramers, December 28, 1923 (AHQP). Kramers' responses, it seems, are no longer extant.} On February 28, 1924, Ladenburg gently reminded Kramers that he had promised in January to give his ``opinion on dispersion and its quantum interpretation"\footnote{Ladenburg to Kramers, February 28, 1924 (AHQP). This fits with Slater's recollection that Kramers already had his new dispersion formula around Christmas 1923.} within a few days. 
A little over a month later, on April 2, Ladenburg wrote another letter to Kramers, in which he thanked him for sending what must have been either a manuscript or proofs of \citep{Kramers 1924a} (which only appeared in the May 10 issue of {\it Nature}) and, apparently having been informed by Kramers that the delay had been due to his wife's illness, apologized for his impatience.\footnote{Ladenburg to Kramers, April 2, 1924 (AHQP). Reiche likewise apologized seven days later (Reiche to Kramers, April 9, 1924 [AHQP]).}

Understandably, given the importance of their own work for Kramers' breakthrough, Ladenburg and Reiche were enthusiastic about the new dispersion formula. Immediately after the one sentence devoted to the illness of Kramers' wife, without so much as starting a new paragraph, Ladenburg wrote in his letter of April 2:
\begin{quotation}
Now your opinion about the dispersion question is of course of the highest interest and I don't want to pass up the opportunity to tell you how much it pleases me that you have managed to give a correspondence derivation of the relation between dispersion and transition probabilities. In this way a solid basis has now been created. Your formula \ldots  is undoubtedly preferable to ours because of its greater generality. I also agree with you that one cannot extract  contributions of the ``negative" oscillators from existing experiments.\footnote{Ladenburg to Kramers, April 2, 1924 (AHQP). Ladenburg was not familiar with the BKS paper at this point, neither with the English version which appeared in April 1924, nor with the German translation which only appeared on May 22.}
\end{quotation}
Ladenburg thus immediately zeroed in on the key experimental question raised by the new formula. In the late 1920s Ladenburg and his collaborators embarked on a ambitious program to verify the second term in the Kramers dispersion formula experimentally. Reiche, writing to Kramers a week later, focused on the theoretical justification of the new formula:
\begin{quotation}
I wanted to tell you again how delighted I am with your beautiful correspondence derivation. Following Epstein's paper [Epstein,1922c] and using the Born-Pauli [1922] method, I easily derived the classical expression for $P$ [the polarization] which you indicate in your letter\footnote{This expression---equivalent eq.\ (\ref{e}) below---is not given in \citep{Kramers 1924a} but does occur in \citep[p.\ 199, eq.\ 2* ]{Kramers 1924b} (reproduced as eq.\ (50) in sec.\ 5.1).} and have also had no trouble reconstructing the correspondence argument for the transition to the quantum formula.\footnote{Reiche to Kramers, April 9, 1924 (AHQP).} 
\end{quotation}  
Fearing that few Germans would have access to {\it Nature}, Ladenburg  and Reiche prepared a detailed report on \citep{Kramers 1924a} for {\it Die Naturwissenschaften}. In late May, Ladenburg asked Kramers whether he would have any objections if they included a derivation of the new dispersion formula, adding that they were not sure how close it was to Kramers' derivation.\footnote{Ladenburg to Kramers, May 31, 1924 (AHQP). Ladenburg and Reiche had meanwhile read \citep{BKS2} and, unsurprisingly given the importance of their concept of `substitute oscillators' for the BKS theory, were instant converts to the theory. For further discussion, see sec.\ 4.} Kramers welcomed the idea, telling Ladenburg that their derivation would probably not be all that different from his own. He had every intention of writing a longer paper on dispersion and absorption himself, he added, which would obviously include the derivation of his dispersion formula, but recognized that ``it will probably be a while before I have time to write such an article; because of lack of time I have not thought through many details and I consequently would not mind it at all if your note appears first."\footnote{Kramers to Ladenburg, June 5, 1924 (AHQP).} 
In the end, the editors of {\it Die Naturwissenschaften} insisted that Ladenburg and Reiche shorten their article.\footnote{Ladenburg to Kramers, June 8, 1924 (AHQP).} It was eventually published without the derivation of Kramers' dispersion formula \citep{Ladenburg and Reiche 1924}.

The distinction of being the first to publish a full derivation of this important result thus fell to Van Vleck. 
Van Vleck read Kramers' first {\it Nature} note not long after he finished his papers on a correspondence principle for absorption \citep{Van Vleck 1924a, Van Vleck 1924b, Van Vleck 1924c}. In a footnote added to the first of these, he wrote:
\begin{quotation}
Since the writing of the present article, Dr.\ H.\ A.\ Kramers has published \ldots a very interesting formula for dispersion, in which the polarization is imagined as coming not from actual orbits, but from ``virtual oscillators" such as have been suggested by Slater and advocated by Bohr. Kramers states that his formula merges asymptotically  [i.e., in the limit of high quantum numbers] into the classical dispersion. To verify this in the general case, the writer has computed the classical polarization formula for an arbitrary non-degenerate multiply periodic orbit \ldots By pairing together positive and negative terms in the Kramers formula, a differential dispersion may be defined resembling the differential absorption of the present article. It is found that this differential quantum theory dispersion approaches asymptotically the classical dispersion \ldots the behavior being very similar to that in the correspondence principle for absorption. This must be regarded as an important argument for the Kramers formula \citep[p.\ 30]{Van Vleck 1924a}.
\end{quotation}
It was not clear to Van Vleck on the basis of Kramers' note exactly what Kramers had and had not yet done. Van Vleck thought that his calculations extended Kramers' results. As he explained to Kramers in September 1924:
\begin{quotation}
[Van Vleck 1924b, c] was ready to send to the printer about the time we received the copy of Nature containing your dispersion formula. In your note [Kramers, 1924a] I did not understand you to state how generally you had verified the asymptotic connection with the classical dispersion from the actual orbit, and it immediately occurred to me that this question could easily be investigated by the perturbation theory method I had previously developed in connection with what I call the ``correspondence principle for absorption".  I therefore inserted two sections (\# 6 and \# 15 \ldots) showing that your formula merged into the classical one. 
Inasmuch as the classical dispersion formula had apparently not been developed for the general non-degenerate multiply periodic orbit, and as you did not give this in your note to Nature, I conjectured that you had verified the asymptotic connection only in special cases, such as a linear oscillator, so that my computations on dispersion would not be a duplication of what you had done. However, while visiting at Cambridge, Mass.\ last week I learned from Dr.\ Slater that your calculation of the asymptotic connection was almost identical with my own in scope and generality. I have therefore altered the proof of my Physical Review article to include a note [Van Vleck 1924b, 345] stating that you have also established the correspondence theorem in the general case.\footnote{Van Vleck to Kramers, September 22, 1924 (AHQP).}
\end{quotation}
As in the case of Ladenburg and Reiche half a year earlier, Kramers did not seem to mind at all that Van Vleck was poaching on his preserves.
He generously wrote back to Van Vleck: ``Your note on absorption made me much pleasure and I think it very just of Providence that you got it published before hearing of our work."\footnote{Kramers to Van Vleck, November 11, 1924 (AHQP).}

The construction of the dispersion formula (\ref{eq:1.4}) requires
as a prelude to the application of the correspondence principle, a derivation of the classical formula for the dipole moment of an arbitrary (non-degenerate) multiply-periodic system. This is where \citet{Ladenburg and Reiche 1923} came up short, even though, as we saw above, Reiche was able to reconstruct the derivation once Kramers had outlined it for him. Kramers and Van Vleck, like Epstein before them, used canonical perturbation techniques from celestial mechanics to derive this classical formula. In Part Two of our paper, closely following the classical part of Van Vleck's two-part paper \citep{Van Vleck 1924c}, we shall present a detailed derivation of this crucial classical formula, for the special case of the harmonic oscillator in sec.\ 5.1 and for a general non-degenerate multiply-periodic system in sec.\ 6.2. Guided by the correspondence principle and introducing the $A$ and $B$ coefficients we then construct a quantum formula that merges with the classical formula for high quantum numbers (see secs.\ 5.1 and 6.2).\footnote{Van Vleck actually does it the other way around: he {\it starts} with the quantum formula and checks that this formula merges with the classical formula in the correspondence limit.} Here we summarize the main steps of this derivation.  

In general coordinates $(q_i, p_i)$, Hamilton's equations are:
\begin{equation}
\label{a}
    \dot{q}_i = \frac{\partial H}{\partial p_i},\;\;\;
    \dot{p}_i = -\frac{\partial H}{\partial q_i},
\end{equation}
where dots indicate time derivatives. Given the Hamiltonian $H$ of some multiply-periodic system, one can often find special coordinates $(w_i, J_i)$, so-called {\it action-angle variables}, in which Hamilton's equations take on a particularly simple form:
\begin{equation}
\label{b}
    \dot{w}_i = \frac{\partial H}{\partial J_i} = \nu_i,\;\;\;
    \dot{J}_i = -\frac{\partial H}{\partial w_i} = 0.
\end{equation}
The angle variables, $w_i = \nu_i t$, give the characteristic frequencies of the system; the (conserved) action variables are subject to the Bohr-Sommerfeld quantum condition, $J_i=n_ih$.

Suppose we have a Hamiltonian $H$ that is the sum of  $H_0$, describing some multiply-periodic system representing an electron orbiting the nucleus of an atom in the Bohr-Sommerfeld theory (or, an inner planet like Mercury orbiting the sun), and  $H_{\rm{int}} = e E x \cos{2 \pi \nu t}$, a small perturbation
describing the interaction of this system with a weak periodic electric field in the $x$-direction (or, the periodic weak gravitational interaction with a distant outer planet). To find the induced polarization responsible for dispersion in this system we need to calculate the coherent part $\Delta x_{\rm coh}$ of the displacement  caused by the perturbation (cf.\ eqs. (\ref{coh})--(\ref{eq:1.1}) in sec.\ 3.1). We assume that the unperturbed system can be solved in action-angle variables, which means that $x(t)$ in the absence of $H_{\rm{int}}$ can be written as a Fourier series:
\begin{equation}
    \label{c}
  x(t) =  \sum_{\tau_i}A_{\tau_i}(J_l)e^{2\pi i \tau_i w_i}.
   \end{equation}
The complex amplitudes have to satisfy the conjugacy relations $A_{\tau_i}=A_{-\tau_i}^{*}$ to ensure that $x(t)$ is real.     
Assuming the interaction is switched on at $t=0$, we can use Hamilton's equations in action-angle variables---still those for $H_0$ rather than those for the full Hamiltonian $H$\footnote{This is where  Van Vleck's calculation differs from those of \citet{Born 1924b} or \citet{Kramers and Heisenberg 1925}.}---to calculate $\Delta w_i$ and $\Delta J_i$ due to the perturbation.  We insert  the results into
 \begin{equation}
  \label{d}
     \Delta x = \sum_k \left( \frac{\partial x}{\partial J_k}\Delta J_k + \frac{\partial x}{\partial w_k}\Delta w_k \right),
     \end{equation}
and collect the coherent terms (i.e., all terms with a factor $e^{2 \pi i \nu t}$). The result is:
 \begin{equation}
 \Delta x_{\rm coh}  = 2eE \sum_{\tau_i} \tau_i \frac{\partial}{\partial J_i}\left(\frac{\tau_i \nu_i}{\nu^{2} - (\tau_i \nu_i)^{2}}|A_{\tau_i}(J_l)|^{2} \right) \cos{2\pi\nu t}.
 \label{e}
   \end{equation}   
For the special case of a charged harmonic oscillator, this expression reduces to the simple expression (\ref{eq:1.1}) found earlier (as we shall show in detail at the end of sec.\ 5.1). 

We now translate this classical formula into a quantum formula. The idea is to construct a quantum formula that merges with the classical formula in the limit of high quantum numbers. This is done in three steps. For high values of the quantum number $i$, the derivatives $\partial/\partial J_i$ can be replaced by difference quotients,\footnote{This replacement is known as ``Born's correspondence rule."  In fact, independently of one another, both Kramers and Van Vleck had found it before Born. We shall return to this point in sec.\ 5.2.} the square of the amplitudes $A_{\tau_i}(J_l)$ by transition probabilities $A_{i \rightarrow j}$ (where $|i -j|$ is small compared to $i$), and orbital frequencies $\nu_i$ by transition frequencies $\nu_{i \rightarrow j}$. We then take the leap of faith that the resulting formula holds for all quantum numbers. Multiplying by the charge $-e$ and the number of atoms $N_i$ to get from the coherent part of the displacement of one atom to the polarization of a group of atoms, we arrive at the Kramers dispersion formula (\ref{eq:1.4}).
     

\subsection{Heisenberg's  {\rm Umdeutung} and dispersion theory}

The Kramers dispersion formula was a crucial step in the transition from the old quantum theory to matrix mechanics, and thereby in the transition from classical phase spaces to Hilbert spaces. As Kramers pointed out in  his second  {\it Nature} note, the formula
\begin{quotation}
only contains such quantities as allow of a direct physical interpretation on the basis of the fundamental postulates of the quantum theory \ldots and exhibits no further reminiscence of the mathematical theory of multiple [sic] periodic systems \citep[p.\ 311]{Kramers 1924b}
\end{quotation}
This point is amplified in the Kramers-Heisenberg paper: 
\begin{quotation}
we shall obtain, quite naturally, formulae which contain only the frequencies and amplitudes {\it which are characteristic for the transitions}, while all those symbols which refer to the mathematical theory of periodic systems will have disappeared \citep[p.\ 234, our emphasis]{Kramers and Heisenberg 1925}.
\end{quotation}
Orbits do not correspond to observable quantities, but transitions do, namely to the frequency $\nu_{i \rightarrow f}$ of the emitted radiation, and, through the Einstein coefficients $A_{i \rightarrow f}$, to its intensity.  In the introduction of his {\it Umdeutung} paper,  \citet{Heisenberg 1925c} explained that he wanted ``to establish a theoretical quantum mechanics, analogous to classical mechanics, but in which only relationships between observable quantities occur" (p.\ 262). In the next sentence he identified the Kramers dispersion theory as one of ``the most important first steps toward such a quantum-theoretical mechanics" (ibid.).  

Rather than using classical mechanics to analyze features of electron orbits and translating the end result into a quantum formula, as Kramers and others had done (cf.\ eqs.\  (\ref{a})--(\ref{e}) above), Heisenberg translated the Fourier series for the position of an electron that forms the starting point of such classical calculations into a quantum expression. He replaced the amplitudes and frequencies by two-index quantities, referring to the initial and final state of a quantum transition, respectively, and thus replaced classical position by an array of numbers associated with transitions between states. Reinterpreting rather than replacing the old theory, he assumed that these new quantities would satisfy all the familiar relations of Newtonian mechanics. Note that Heisenberg thus formulated a new theory directly in terms of transition quantities without bothering to find a representation for the states connected by the transitions.

The Bohr-Sommerfeld  quantization condition (\ref{eq:1.0}) has the form of a restriction on orbits in phase space. With the elimination of orbits, it could no longer be used, at least not in its original form. 
As Heisenberg recalls in his AHQP interview:
\begin{quotation}
I had, of course, to think about the quantum condition. And that was an important point. But there I knew so much from Copenhagen how important this Thomas-Kuhn sum rule was. That took some time. That I think I had done in G\"ottingen, [I] had seen how I could translate the Thomas-Kuhn sum rule into what I call a quantum mechanical statement, into a statement in which only differences occurred. I did not see that it was a commutation rule [but with this translation] I can bring this sum rule into my whole scheme and then this sum rule actually fixes everything. I could see that this fixes the quantization.\footnote{P.\ 10 of the transcript of session 7 of the AHQP interview with Heisenberg} 
\end{quotation}
The Thomas-Kuhn sum rule, a corrolary of the Kramers dispersion formula (see sec.\ 7.1 a derivation in modern quantum mechanics), had been found independently  by Werner Kuhn (1899--1963) (1925) in Copenhagen\footnote{The publication of Kuhn's paper had been delayed in typical Copenhagen fashion: ``A paper on the summation rule had been submitted to Prof.\ Bohr and Prof.\ Kramers about half a year before the final one, but it was rejected at that time because it contained besides the main good argument some unsuitable passages" (Werner Kuhn to Thomas Kuhn, May 3, 1962 [included in the folder on Kuhn in the AHQP])} and by Willy Thomas (1925) in Breslau.\footnote{Thomas was a student of Reiche in Breslau who died young of tuberculosis. See p.\ 14 of the transcript of the third session of the AHQP interview with Reiche.} Kuhn (i.e., Thomas Kuhn) pressed Heisenberg a little on how he had settled on this rule as his fundamental quantization condition: ``Using the Kuhn-Thomas [sic] rule is a stroke of genius but one supposes that there were a lot of other intermediate attempts." Apparently there were none. Heisenberg insisted:
\begin{quotation}
No, I would say it was rather trivial for the following reasons: First of all, there was the integral pdq. Then one had seen that integral pdq sometimes is 1/2 and sometimes is not 1/2. That played a role. Because then I felt that perhaps only the difference of integral pdq between one quantum state and the next quantum state is an important thing. So I actually felt, ``Well, perhaps I should write down integral pdq in one state minus integral pdq in the neighboring state." Then I saw that if I write down this and try to translate it according to {\it the scheme of the dispersion theory}, then I get the Thomas-Kuhn sum rule. And that is the point. Then I thought, ``Well, that is apparently the way how it is done."\footnote{P.\ 10 of the transcript of session 7 of the AHQP interview with Heisenberg (our emphasis).} 
\end{quotation}
In other words, following the general recipe introduced in the {\it Umdeutung} paper for the translation of classical formulae into quantum-mechanical ones--- ``the scheme of the dispersion theory"---\citet[p.\ 268]{Heisenberg 1925c} was able to convert {\it a derivative of} the Bohr-Sommerfeld condition into an equation that contains only amplitudes and frequencies. Since Heisenberg's theory only deals with transitions between states, the absolute value of the action $J$ does not matter; only the difference in $J$-value between two states does.\footnote{The problem with half-integer values for $J$ that Heisenberg mentioned in the passage from the AHQP interview quoted above is not mentioned in the {\it Umdeutung} paper.}

The sum rule is sometimes called the Thomas-Kuhn-Reiche sum rule because \citet{Reiche and Thomas 1925} were the first to publish a detailed derivation of it in  a paper  {\it submitted} to the {\it Zeitschrift f\"{u}r Physik} in early August 1925 about a month before \citep{Heisenberg 1925c} {\it appeared} in the same journal. In formulating the goal of their paper, Reiche and Thomas not only used the term `Umdeutung' in very much the same way as Heisenberg in his {\it Umdeutung} paper, they also explicitly tied this usage to Kramers' dispersion theory: 
\begin{quotation}
We use \ldots the correspondence principle in the same way in which it was applied by Kramers in the derivation of the dispersion formula by reinterpreting ({\it umdeuten}) the mechanical orbital frequencies as radiation frequencies, the Fourier coefficients as the ``characteristic amplitudes" that determine the quantum radiation, and, finally, in analogy to the Bohr frequency condition,\footnote{In the limit of high quantum numbers, the Bohr frequency condition, $\nu_{i \rightarrow j} = (E_i - E_j)/h$, merges with the relation $\nu_i = \partial H/\partial J_i$ (cf.\ eq.\ (\ref{b})). \citet[p.\ 333]{Van Vleck 1924b} calls this the correspondence theorem for frequencies.} differential quotients as difference quotients. In the realm of high quantum numbers the classical and quantum-theoretical representations become identical. We try to arrive at a general relation, by maintaining the reinterpretation ({\it Umdeutung}) of classical quantities into quantum-theoretical ones for all quantum numbers  \citep[pp.\ 511--512]{Reiche and Thomas 1925}.
\end{quotation}
In view of the tendency of European theorists to neglect American contributions (see sec.\ 2.4), it is also interesting to note that \citet[p.\ 513]{Reiche and Thomas 1925} cite \citep{Van Vleck 1924b}.

Although he failed to recognize the importance of the result at the time, Van Vleck had, in fact, been the first to find the sum rule \citep[p.\ 135, note 184]{Sopka}. As he wrote in his NRC {\it Bulletin}:
\begin{quotation}
Eq.\ (62a) [a version of the sum rule] appears to have been first incidentally suggested by the writer [Van Vleck 1924c, pp. 359--360, footnote 43] and then was later and independently much more strongly advanced by Thomas \ldots Kuhn \ldots and Reiche and Thomas \citep[p.\ 152]{Van Vleck 1926a}.
\end{quotation}
Van Vleck is referring to a footnote in the section on dispersion in the classical part of his paper. In this footnote he mentioned two objections that explain why he did not put greater emphasis on the sum rule himself. Van Vleck's idea---which he calls ``tempting (but probably futile)" \citep[p.\ 359, footnote 43]{Van Vleck 1924c}---was that the sum rule would allow him to compute the Einstein $A$ coefficients. He was under the impression, however, that ``such a method is hard to reconcile with the [experimental] work of F.\ C.\ Hoyt [1923, 1924]" on X-ray absorption and that it ``would lead to transitions from positive to negative quantum numbers, which can scarcely correspond to any physical reality" (ibid.).     

As \citet[pp.\ 269--270]{Heisenberg 1925c} shows briefly in his paper, the sum rule follows from the Kramers dispersion formula (\ref{eq:1.4}) if one takes the limit in which the frequency $\nu$ of the incident radiation is much greater than any of the absorption frequencies $\nu_{i \rightarrow j}$ (see sec.\  7.1). That the quantization condition obtained by massaging the Bohr-Sommerfeld condition also follows from the Kramers dispersion theory, widely  recognized as one of the most secure parts of the old quantum theory, clearly bolstered Heisenberg's confidence in the translation procedure of his {\it Umdeutung} paper. 
It was left to \citet{Born and Jordan 1925b} to extract  the now standard commutation relations for position and momentum from the Thomas-Kuhn sum rule (in sec.\ 7.1 we shall show in detail how this is done). That Heisenberg stopped short of making this move, as we shall argue in sec.\  7.1, is largely because he was thinking in terms of the positions and velocities of the Lagrangian formalism rather than in terms of the positions and momenta of the Hamiltonian formalism.

Although Heisenberg thus relied heavily on dispersion theory in his {\it Umdeutung} paper, he gave his positivist methodology pride of place. His positivism probably came from a variety of sources. Pauli, Heisenberg's fellow student and frequent discussion partner (both in person and in writing), was a devoted follower of his godfather Ernst Mach (1838--1916).\footnote{On Pauli's positivism, see, e.g., \citep[pp.\ 19--23]{Hendry 1984} and \citep{Gustavson 2004}.} As Pauli had written to Bohr, for instance, on December 12, 1924:
\begin{quotation}
We must not \ldots put the atoms in the shackles of our prejudices (of which in my opinion the assumption of the existence of electron orbits in the sense of the ordinary kinematics is an example); on the contrary, we must adapt our concepts to experience \citep[Vol.\ 5, pp.\ 35--36]{Bohr 1972--1996}. 
\end{quotation}
We already mentioned in sec.\ 1.1 that Heisenberg himself later claimed that his positivist attitude came in part from his reading of Einstein's 1905 special relativity paper.\footnote{See, e.g., \citep[pp.\ 26--31]{Holton 2005} for discussion.} His biographer David \citet[p.\ 198]{Cassidy} makes the suggestive observation that  \citet[p.\ 493]{Born and Jordan 1925a}, in a paper completed by June 11, 1925, not only emphasized the observability principle but also appealed to Einstein's analysis of distant simultaneity in support of it.      

As Helge \citet{Kragh 1999} notes: ``there was no royal road from the observability principle to quantum mechanics" (p.\ 162). This truism is nicely illustrated by 
a conversation between Einstein and Heisenberg reported years later by the latter. The following exchange supposedly took place in Berlin in the spring of 1926:
\begin{quotation}
``But you don't seriously believe," Einstein protested, ``that none but observable magnitudes must go into a physical theory?" 
``Isn't that precisely what you have done with relativity?" I asked in some surprise \ldots ``Possibly I did use this kind of reasoning," Einstein admitted, ``but it is nonsense all the same" \citep[p.\ 63]{Heisenberg 1971}.\footnote{Quoted and discussed, for instance, in \citep[p.\ 185]{MacKinnon 1977} and in \citep[pp.\ 30--31]{Holton 2005}. For other versions of the same anecdote, see \citep[pp.\ 113--114]{Heisenberg 1983} and pp.\ 18--19 of the transcript of session 5 of the AHQP interview with Heisenberg.\label{WH5}} 
\end{quotation}
With his $S$-matrix program in the 1940s,\footnote{See \citep[497--503]{Pais 1986}, \citep[453--458]{Dresden}, and, especially, \citep{Cushing 1990} for discussion. See also pp.\ 20--21 of the transcript cited in note \ref{WH5}.} 
Heisenberg once again tried to force a theoretical breakthrough by restricting himself to observable quantities, this time with the qualification that he had taken to heart Einstein's lesson that, in the end, it is the theory that determines what the observables are. \citet[p.\ 63]{Heisenberg 1971} has Einstein make this point a few sentences after the passage quoted above and acknowledges it as a source of inspiration for his 1927 uncertainty principle. Nearly two decades after the {\it Umdeutung} paper, \citet{Heisenberg 1943} wrote: ``in this situation it seems useful to raise the question which concepts of the present theory can be retained in the future theory, and this question is roughly equivalent to a different question, namely which quantities of the current theory are ``observable" \ldots Of course, it will always only be decided by the completed theory which quantities are truly ``observable"" (p.\ 514).

As Einstein complained in 1917 in a letter to his friend Michele Besso (1873--1955), referring to  the excessive Machian positivism of their mutual acquaintance Friedrich Adler (1879--1960): ``He is riding the Machian nag [{\it den Machschen Klepper}] to exhaustion." In a follow-up letter he elaborated: ``It cannot give birth to anything living, it can only stamp out harmful vermin."\footnote{Einstein to Besso, April 29 and May 13, 1917, respectively \citep[Vol.\ 8, Docs.\ 331 and 339]{Einstein 1987--2004}. For further discussion, see, e.g., \citep{Holton 1968}.} This is true in the case of matrix mechanics as well.  Heisenberg's positivism would have been perfectly sterile if it had not been for Kramers' dispersion theory. In that context, positivism was not a blanket injunction against unobservable quantities in general but was directed at a specific set of increasingly problematic unobservables, the electron orbits of the Bohr-Sommerfeld theory.  


\section{The Bohr-Kramers-Slater (BKS) theory as a detour on the road from dispersion theory to matrix mechanics}

\subsection{Virtual oscillators and virtual radiation}

Kramers presented his work on dispersion theory in the context of the BKS theory, not just in the two preliminary notes to {\it Nature} discussed in sec.\ 3.4, but also in the authoritative exposition of his dispersion theory in the paper with Heisenberg.
 In the abstract of this paper, the authors announce that
\begin{quotation}
[t]he arguments are based throughout on the interpretation of the connection of the wave radiation of the atom with the stationary states advocated in a recent paper by Bohr, Kramers and Slater [1924a,b], and the conclusions, should they be confirmed, would form an interesting support for this interpretation \citep[p.\ 223]{Kramers and Heisenberg 1925}.\footnote{We use the translation of \citet[p.\ 87]{Stolzenburg 1984} at this point, which is more accurate than the standard translation in \citep[p.\ 223]{Van der Waerden}.}
\end{quotation}
It should thus come as no surprise that the Kramers dispersion theory has been portrayed as an application of the BKS theory in most  older and even in some more recent historical literature.\footnote{There is an extensive literature on the BKS theory; see, e.g., \citep[pp.\ 23--39]{Klein 1970}, \citep[pp.\ 291--305]{Stuewer 1975}, \citep{Hendry 1981}, the dissertation of Neil \citet{Wasserman}, \citep[Vol.\ 1, sec.\ V.2]{Mehra Rechenberg}, the essay by Klaus \citet{Stolzenburg 1984} in \citep[Vol.\ 5, pp.\ 3--96]{Bohr 1972--1996}, and \citep[pp.\ 159--215]{Dresden}.}   Max \citet{Jammer 1966}, for instance, writes that the BKS theory ``was the point of departure of Kramers's detailed theory of dispersion" (p.\ 184). Mara \citet{Beller} still characterized \citep{Kramers and Heisenberg 1925} as a paper that ``spelled out, in a rigorous mathematical way, the ideas only roughly outlined in the presentation of Bohr, Kramers, and Slater" (p.\ 23). More than a decade earlier, \citet[pp.\ 144--146, pp.\ 220--221]{Dresden} had in fact already set the record straight.\footnote{See \citep[p.\ 221]{Dresden} for a helpful chronology of events in 1923--1925 pertaining to BKS and dispersion theory.} \citet[p.\ 225]{Darrigol} duly emphasizes that {\it the Kramers dispersion theory was developed before and independently of the BKS theory}.  Even before Dresden, \citet{Hendry 1981} had already made it clear that {\it BKS got its virtual oscillators from dispersion theory---the substitute oscillators of \citep{Ladenburg and Reiche 1923}---and not the other way around}. We briefly review the evidence in support of the italicized claims above.

We know from the passage quoted in sec.\  3.4 from a letter from Slater to Van Vleck that by the time the former arrived in Copenhagen around Christmas 1923 Kramers already had his dispersion formula. Kramers must have used the substitute oscillators of \citet{Ladenburg and Reiche 1923} at that point even though by the time he finally got around to publishing his formula he called them virtual oscillators (see sec.\ 3.4). Slater's arrival in Copenhagen marks the lower limit for the birth of the BKS theory.  The theory, after all, grew around an idea that Slater hit upon shortly before he left for Europe late that year.\footnote{See Slater to his mother, November 8, 1923 (quoted in Dresden, 1987, p.\ 161); Slater to Kramers, December 8, 1923 (AHQP). For discussions of Slater's idea, see \citep[p.\ 23]{Klein 1970}, \citep[pp.\ 291--294]{Stuewer 1975}, \citep[pp.\ 213--214]{Hendry 1981}, \citep[pp.\ 6--11]{Stolzenburg 1984}, and \citep[pp.\ 218--219]{Darrigol}.} Slater suggested that the wave and particle properties of light might be reconciled by having an  electromagnetic field guide corpuscular light quanta.\footnote{Slater was probably unaware that Einstein and Louis de Broglie (1892--1987) had already made similar suggestions (Hendry, 1981, p.\ 199; Darrigol, 1992, p.\ 218).} Bohr and Kramers cannibalized Slater's  idea and  stripped it of all reference to light quanta. Against his better judgment---as he insisted decades later in a letter of November 4, 1964 to van der Waerden (1968, p.\ 13)---Slater went along and his idea entered the literature via the BKS paper. In a short letter sent to {\it Nature} a week after this joint paper had been submitted, Slater explained how Bohr and Kramers had convinced him of their point of view. Accordingly, he presented his idea couched in BKS terms:
\begin{quotation}
Any atom may, in fact, be supposed to communicate with other atoms all the time it is in a stationary state, by means of a virtual field of radiation originating from oscillators having the frequencies of possible quantum transitions and the function of which is to provide for the statistical conservation of energy and momentum by determining the probabilities for quantum transitions \citep[p.\ 307]{Slater 1924}.
\end{quotation}
The final clause about the statistical conservation of energy and momentum was foisted upon Slater by Bohr and Kramers.\footnote{See also \citep[p.\ 160]{BKS}.} Bohr had been contemplating such a move for several years, as can be inferred, for instance, from correspondence with Ehrenfest in 1921 in connection with the third Solvay congress held that year \citep[p.\ 19]{Klein 1970} and with Darwin in 1922 \citep[pp.\ 13--19]{Stolzenburg 1984}.
Slater's concept of virtual radiation emitted while an atom is in a stationary state fit nicely  with Bohr's tentative ideas concerning the mechanism of emission and absorption of radiation. In secs.\ 3.2--3.3, we quoted various comments by Bohr on dispersion  from the period 1916--1923 showing how he came to embrace the notion that an atom interacts with radiation like a set of oscillators. 

The concept of virtual oscillators is often attributed to Slater, not just by later historians (see, e.g., Stuewer, 1975, p.\ 291, p.\ 303) but also by his contemporaries. 
In the abstract of  \citep{Van Vleck 1924b}, for instance, we read that the Kramers dispersion formula  ``assumes the dispersion to be due not to the actual orbits but to Slater's `virtual' or `ghost' oscillators having the spectroscopic rather than orbital frequencies" (p.\ 330).\footnote{See also \citep[p.\ 30]{Van Vleck 1924a}, quoted in sec.\ 3.4, and \citep[p.\ 163]{Van Vleck 1926a}.} 
In the BKS paper itself, however, the concept is unambiguously attributed to Ladenburg:\footnote{The mistakes with the prepositions in the passage below (`reaction on' instead of `reaction to' and `considerations on' instead of `considerations of') would tend to support Slater's claim that the paper was ``written entirely by Bohr and Kramers" (Slater to Van Vleck, July 27, 1924, quoted in sec.\  2.2).}
\begin{quotation}
The correspondence principle has led to comparing the reaction of an atom on a field of radiation with the reaction on such a field which, according to the classical theory of electrodynamics, should be expected from a set of `virtual' harmonic oscillators with frequencies equal to those determined by [$h\nu = E_1 - E_2$] for the various possible transitions between stationary states.\footnote{At this point, the authors refer to Ch.\ III, sec.\ 3 of the (English translation of) \citep{Bohr 1923b}, the section in which \citep{Ladenburg 1921} is discussed and which triggered the correspondence between Bohr and Ladenburg  discussed in sec.\ 3.3.} Such a picture has been used by Ladenburg\footnote{At this point, the authors append a footnote referring to \citep{Ladenburg 1921} and \citep{Ladenburg and Reiche 1923}.} in an attempt to connect the experimental results on dispersion quantitatively with considerations on the probability of transitions between stationary states \citep[pp.\ 163--164]{BKS}.
\end{quotation}
As we saw in sec.\ 3.3, Ladenburg and Reiche in turn attributed the idea to Bohr. In 1924, for instance, they wrote:
\begin{quotation}
Formally, we can describe the relation [between oscillator strengths and transition probabilities] following an assumption introduced by Bohr [1923b, pp.\ 161--162], by imagining that the atom responds to external radiation like a system of electrical oscilators, whose characteristic frequencies $\nu$ agree with the emitted or absorbed frequencies in {\it possible} quantum transitions \citep[p.\ 672]{Ladenburg and Reiche 1924}.
\end{quotation}
In the next sentence they use their own term ``substitute oscillators" (in quotation marks) and add: ``(now called ``virtual oscillators")" (ibid.). Likewise, in the introduction of the opening installment of a series of papers on the experimental verification of the Kramers dispersion formula, Ladenburg talks about ``the ``substitute oscillators"," which were introduced, ``at Bohr's suggestion, as the carriers of the scattered radiation needed for dispersion" \citep[p.\ 16]{Ladenburg 1928}.\footnote{The passage from which these clauses are taken is quoted in full at the beginning of sec.\ 7.}

Bohr had communicated the idea in a letter to Ladenburg (see sec.\ 3.3). This may explain why, when interviewed for the AHQP, Reiche did not remember who originally came up with it:
\begin{quotation}
I do not know whether we or Kramers first used this terminology of virtual oscillators \ldots It might be it is Kramers. If it was Kramers then we certainly at once incorporated it into our thinking.\footnote{See p.\ 11 of the transcript of  session 3 of the interview with Reiche. It could be, however, that Reiche was only referring to the new {\it term} for the Bohr-Ladenburg-Reiche concept of substitute oscillators.}
\end{quotation}
In his AHQP interview with Slater, Kuhn also asked about virtual oscillators: 
\begin{quotation}
to what extent did that come from [the BKS] paper, to what extent does it really go back to the Ladenburg, and Ladenburg-Reiche? It could have grown out [of the] Ladenburg and Ladenburg-Reiche papers, yet my impression from the literature is that there was little done with that until after  the Bohr-Kramers-Slater paper.\footnote{P.\ 34 of the transcript of the first session of the AHQP interview with Slater.}
\end{quotation}
Slater concurred, though his comments would have been more valuable had he not been asked such a leading question:
\begin{quotation}
I think that's true. Of course, I was very familiar with the Ladenburg-Reiche things,\footnote{\citep{Ladenburg 1921} and \citep{Ladenburg and Reiche 1923} are cited in \citep[p.\ 397]{Slater 1925a}.} so was Bohr. I think that we helped popularize it in a sense. Of course, this also came at the same time, approximately, that Kramers was working on his dispersion formula. That again is operating with things very much like the virtual oscillator, so they all seem to hang together, and I think it was a combination of the oscillators from our paper, from the Ladenburg-Reiche, and the Heisenberg-Kramers dispersion that really set them in operation.\footnote{Ibid., pp.\ 34--35.}
\end{quotation}
Despite the loaded question that elicited this response and even though Slater is wrong to suggest that  BKS and Kramers' dispersion theory were developed independently of the earlier work of Ladenburg and Reiche, the overall characterization of the situation seems to be accurate.  BKS officially sanctioned the dual representation of the atom as simultaneously a quantum system \`{a} la Einstein and Bohr and a set of oscillators \`{a} la Lorentz and Drude. This dual picture had been implicit in \citep{Ladenburg 1921}. It was made explicit, under Bohr's influence, in \citep{Ladenburg and Reiche 1923}. That it was endorsed by the highest authorities in Copenhagen undoubtedly helped its dissemination. Even so it was typically presented with some trepidation. In his second {\it Nature} note, Kramers passed it off as merely a matter of words:\footnote{In the work that led to \citep{Kramers and Heisenberg 1925}, however, Kramers, according to \citet{Hendry 1981}, ``ignored their virtual nature altogether and treated the oscillator model as naively as he had the orbital model" (p.\ 202).} 
\begin{quotation}
In this connexion it may be emphasized that the notation `virtual oscillator' used in my former letter [Kramers, 1924a] does not mean the introduction of any additional hypothetical mechanism, but is meant only as a terminology suitable to characterise certain main features of the connexion between the description of optical phenomena and the theoretical interpretation of spectra \citep[p.\ 311]{Kramers 1924b}.
\end{quotation}
Van Vleck was more upfront:
\begin{quotation}
The introduction of these virtual resonators is, to be sure, in some ways very artificial, but is nevertheless apparently the most satisfactory way of combining the elements of truth in both the classical and quantum theories. In particular this avoids the otherwise almost insuperable difficulty that it is the spectroscopic rather than the orbital frequencies \ldots which figure in dispersion \citep[p.\ 344]{Van Vleck 1924b}.
\end{quotation}
Despite such disclaimers, Kramers and Van Vleck---as well as Slater, Born, Breit and others working in the general area of dispersion theory in 1924--1925---used a model of the atom in which the electron orbits of the Bohr-Sommerfeld theory were supplemented by an ``orchestra of virtual oscillators"\footnote{The term ``virtual orchestra" comes from \citep[p.\ 456]{Lande 1926} \citep[p.\ 187]{Jammer 1966}.} with characteristic frequencies corresponding to each and every transition that an electron in a given orbit can undergo. Thanks to virtual oscilla\-tors---to paraphrase Heisenberg's succinct statement to van der Waerden (1968, p.\ 29) in 1963---at least {\it something} in the atom was vibrating with the right frequency again.

The dual representation of physical systems (of electrons rather than atoms in this case) was also key to the BKS explanation of the Compton effect. BKS was Bohr's last stand against light quanta
after the Compton effect had finally convinced most other physicists that they were unavoidable \citep[p.\ 3]{Klein 1970}.\footnote{Cf.\ our comments in the introduction to sec.\ 3.} BKS explains the Compton effect without light quanta. It attributes the frequency shift between incoming and scattered X-rays to a Doppler shift in the X-ray wave fronts instead.  \citet{Compton 1923} thought this option was ruled out because, as he showed in his paper, the recoil velocity needed to get the right Doppler shift is different from the recoil velocity needed to ensure conservation of energy and momentum in the process, and one and the same electron cannot recoil with two different velocities. In the BKS theory, however, there is room for two recoil velocities, one for the electron itself, one for the orchestra of virtual oscillators associated with it.\footnote{What \citet{Compton 1923} actually said in his paper is very suggestive of this option: ``It is clear \ldots that so far as  the effect on the wave-length is concerned, we may replace the recoiling electron by a scattering electron" with an ``effective velocity" different from that of the recoiling electron (p.\ 487; quoted and discussed in Stuewer, 1975, p.\ 230).} The Compton effect can be interpreted as a Doppler shift if the appropriate recoil velocity is assigned to the virtual oscillators.  Energy and momentum can be conserved if a different recoil velocity is assigned to the electrons themselves. Bohr and his co-authors wasted few words on the justification of this startling maneuver:
\begin{quotation}
That in this case the virtual oscillator moves with a velocity different from that of the illuminated electrons themselves is certainly a feature strikingly unfamiliar to the classical conceptions. In view of the fundamental departures from the classical space-time description, involved in the very idea of virtual oscillators, it seems at the present state of science hardly justifiable to reject a formal interpretation as that under consideration as inadequate \cite[p.\ 173]{BKS}.
\end{quotation}
This is almost as bad as pieces of glass dragging along different amounts of ether for different colors of light in early-19th-century ether theory (see sec.\ 3.1)! 

The problem carries over to the dispersion theory based on the dual representation of atoms in terms of classical orbits and virtual oscillators, as is acknowledged, if only in passing, by 
\citet{Kramers and Heisenberg 1925}: ``We shall not discuss in any detail the curious fact that the centre of these spherical waves moves relative to the excited atom" (p.\ 229). This exacerbated the problem of the Bohr-Sommerfeld orbits in the theory. Not only were they responsible for the discrepancy between orbital frequencies and radiation frequencies, they also make it harder to picture an atom in space and time.  After all, the system of electron orbits does not even move in concert with its orchestra of virtual oscillators. 

Edward \citet{MacKinnon 1977, MacKinnon 1982} has suggested that the resulting problem of combining different pictures of the atom into one coherent picture forced Heisenberg to make a choice between them (see also Beller, 1999, p.\ 23). Since the virtual oscillators carry all the physical information while the electron orbits are completely unobservable, the choice is obvious. \citet[p.\ 138]{MacKinnon 1977} has gone as far as describing Heisenberg's {\it Umdeutung} paper as proposing a theory of virtual oscillators. Of course, there is no explicit reference to virtual oscillators anywhere in the {\it Umdeutung} paper. \citet[pp.\ 155--156, 162, 177]{MacKinnon 1977} speculates that this is because Heisenberg suppressed all talk about virtual oscillators as a response to Pauli's objections to the ``virtualization" of physics.\footnote{In a letter of January 8, 1925, Heisenberg told Bohr that Pauli did not believe ``in virtual oscillators and is outraged at the `virtualization' of physics" \citep[p.\ 156]{MacKinnon 1977}.}  We shall return to the relation between BKS and Heisenberg's work in sec.\ 4.3. 

Pauli had originally promised not to subvert Bohr's efforts to get the physics community to accept the term `virtual'  as used in the context of BKS.  Working on the German translation of the paper \citep{BKS2}, Bohr was anxious to ensure that Pauli approved of ``the words ``communicate" and ``virtual", for after lengthy consideration, we have agreed here on these basic pillars of the exposition."\footnote{Bohr to Pauli, February 16, 1924 \citep[Vol.\ 5, p.\ 409]{Bohr 1972--1996}.} 
In typical Bohr fashion, he first announced that the manuscript would be submitted that same day 
and that he would enclose a copy, then added a postscript saying that there had been further delays and that it would be sent later.\footnote{Contrary to what is suggested by these delays, the German translation simply follows the English original.} Amused, Pauli wrote back a few days later:
\begin{quotation}
I laughed a little (you will certainly forgive me for that) about your warm recommendation of the words ``communicate" and ``virtual" and about your postscript that the manuscript is still not yet completed. On the basis of my knowledge of these two words ({\it which I definitely promise you not to undermine}), I have tried to guess what your paper may deal with. But I have not succeeded.\footnote{Pauli to Bohr, February 21, 1924 \citep[Vol.\ 5, p.\ 412; our emphasis]{Bohr 1972--1996}. Our reading of this letter differs from that of \citet{Hendry 1981}, who characterized it as ``mocking" (p.\ 202).}
\end{quotation}
The term `virtual' also puzzled the group of physicists in Ann Arbor studying the BKS paper with Bohr's former associate Klein, who wrote to Bohr on June 30, 1924: ``Colby [cf.\ note \ref{colby}], who is also most interested in it, asked me about the meaning of the term `virtual radiation'" \citep[p.\ 29]{Stolzenburg 1984}. 

Exactly what does the `virtual' in virtual oscillator and virtual radiation mean? Virtual oscillators can be thought of in analogy to virtual images in geometric optics. Just as the light reflected from a mirror appears to come from an imaginary point behind the mirror, the light scattered by an atom appears to come from an imaginary oscillator. This analogy, however, is nowhere to be found in the BKS paper. Whatever its exact meaning, the designation `virtual' does serve as a warning that these oscillators are not just classical oscillators. The authors warn, for instance, that ``the absorption and emission of radiation are coupled to different processes of transition, and thereby to different virtual oscillators" \citep[p.\ 171]{BKS}. The virtual oscillators are, as \citet{Darrigol} summarizes the situation, ``nothing but a condensed expression of their effects, which could be deduced from the correspondence principle piece by piece but could not be synthesized in any classical model" (p.\ 257). 

Unlike the light coming from virtual images in geometric optics, the radiation coming from virtual oscillators  is {\it also} called virtual in the BKS paper. Again, it is not exactly clear why. As the analogy with geometric optics shows, that a source is virtual does not necessarily mean that the radiation is virtual as well. In Slater's original conception, the radiation might be called virtual in the sense that the light quanta are the primary reality  and that the radiation is there only to guide them. In the BKS theory, however, there are no light quanta, only the radiation. 

The way Heisenberg later remembered it, the virtual radiation of the BKS theory had the same status as the Schr\"odinger wave function in Born's statistical interpretation a few years later. As Heisenberg told Kuhn in his AHQP interview:
\begin{quotation}
everybody felt that paper [BKS] contained an essential part of truth \ldots What Bohr, Kramers, and Slater did was to establish the probability as a kind of reality \ldots one felt that by making the probability become some kind of reality, you get hold of something which is there. It was at that time of course, very difficult to say what it was that you had gotten hold of. I would say only through the paper of Born [1926] did it become quite clear that one should say, ``All right, the Schr\"odinger wave means that probability that an electron should be there." But the main point was that the probability itself was something real. It was not only in the mind of the people, but it was something in nature \ldots Up to that time people had two possibilities. One possibility was that the reality is a wave. There is an electric field, and a magnetic field acting upon an atom, shaking the electron, and then the atom does something, it makes a transition \ldots There is an entirely different picture of reality in which there is a light quantum \ldots hitting the atom, and then something happens. But now the idea is that there is a wave. But this wave is not the reality. This wave is a probability---this wave is a tendency. It means that when this wave is present then the atom gets a tendency to emit light quanta. So this idea of the wave field being a tendency was something just in the middle between reality and non-reality \ldots That was the striking thing about it, you know, this new invention of a possibility which was a reality in some way but not a real reality---a half reality.\footnote{P.\ 2 of the transcript of session 4 of the AHQP interview with Heisenberg}
\end{quotation}
Unsurprisingly, Born took exception to Heisenberg's suggestion that the Born interpretation had been anticipated in this way by BKS. As Heisenberg said in a subsequent session of the interview: ``I felt once, when I discussed this matter with Born, that he was a bit angry that I had quoted too much the Bohr-Kramers-Slater paper in connection with the probability interpretation of waves."\footnote{P.\ 21 of the transcript of session 6 of the AHQP interview with Heisenberg.} We sympathize with Born. Heisenberg's comments, we feel, have all the flavor of an after-the-fact rationalization.

In subsequent expositions of the BKS theory by both Kramers and Slater, the radiation from virtual oscillators is presented as every bit as real as the external radiation. It is hard to see how this could be otherwise since the two types of radiation are supposed to interfere with one another. \citet{BKS} write:  ``we shall assume that [illuminated atoms] will act as secondary sources of {\it virtual} wave radiation which interferes with the incident radiation" (p.\ 167, our emphasis). A few pages later, they talk about the same ``secondary wavelets set up by each of the illuminated atoms" (ibid., p.\ 172) without labeling them virtual. On the following page they suddenly refer to the external radiation as ``incident {\it virtual} radiation" (ibid., p.\ 173, our emphasis). And the final paragraph of the paper discusses the  ``(virtual) radiation field" (ibid., p.\ 175) produced by ordinary antennas. The concluding sentence, which has Bohr written all over it, shows how the authors struggled with their own terminology:
\begin{quotation}
It will in this connexion be observed that the emphasizing of the `virtual' character of the radiation field, which at the present state of science seems so essential for an adequate description of atomic phenomena, automatically loses its importance in a limiting case like that just considered [i.e., a classical antenna], where the field, as regards its observable interaction with matter, is endowed with all the attributes of an electromagnetic field in classical electrodynamics \citep[p.\ 175]{BKS}.
\end{quotation}
Subsequent expositions of the BKS theory by Slater and Kramers removed much of the tentativeness of this passage.

In a lengthy paper signed December 1, 1924, and published in the April 1925 issue of {\it The Physical Review}, Slater tried to work out a ``consistent detailed theory of optical phenomena" based on the BKS theory \citep[p.\ 395]{Slater 1925a}. Slater presented this work at a meeting of the {\it American Physical Society} in Washington, D.C., in December 1924 \citep{Slater 1925b}. At this same meeting---which also marked the end of the controversy between Compton and Harvard's William Duane (1872--1935) over the Compton effect \citep[p.\ 273]{Stuewer 1975}---\citet{Van Vleck 1925} talked about \citep{Van Vleck 1924b, Van Vleck 1924c} and \citet{Breit 1925a} talked about \citep{Breit 1924a}.\footnote{The AHQP contains some correspondence between Slater and Van Vleck regarding this meeting and regarding \citep{Slater 1925a}: Slater to Van Vleck, December 8, 1924; Van Vleck to Slater, December 15, 1924.} Slater sent a copy of his paper to Bohr in December 1924 and defended his elaboration of the BKS theory in a letter to Bohr of January 6, 1925 \citep[Vol.\ 5, pp.\ 65--66]{Bohr 1972--1996}. 

In the introduction of his paper, Slater presents the dilemma that led him to embrace Bohr's statistical conservation laws.\footnote{See also the brief discussion of the BKS theory in \citep[pp.\ 285--286]{Van Vleck 1926a}.} The problem, he argues, is that
\begin{quotation}
in the quantum theory the energy of atoms must change by jumps; and in the electromagnetic theory the energy of a radiation field must change continuously \ldots Two paths of escape from this difficulty have been followed with more or less success. The first is to redefine energy [i.e., to adopt Einstein's light-quantum hypothesis]; the second to discard conservation. Optical theory on [the first interpretation] would be a set of laws telling in what paths the quanta travel \ldots [One way to do this is] to set up a sort of ghost field, similar to the classical field, whose function was in some way to guide the quanta. For example, the quanta might travel in the direction of Poynting's vector in such a field. The author was at one time of the opinion that this method was the most hopeful one for solving the problem \ldots The other direction of escape from the conflict between quantum theory and wave theory has been to retain intact the quantum theory and as much of the wave theory as relates to the field, but to discard conservation of energy in the interaction between them \citep[pp.\ 396--397]{Slater 1925a}.
\end{quotation}
Slater sketches some difficulties facing this second approach, but makes it clear that this is the approach he now favors:
\begin{quotation}
An attempt was made by the writer, in a note to Nature [Slater, 1924], enlarged upon in collaboration with Bohr and Kramers, to contribute slightly to the solution of these difficulties. In the present paper, the suggestions made in those papers are developed into a more specific theory (ibid., p.\ 398). 
\end{quotation}
Slater then  describes more carefully how to picture the interaction between matter and radiation in the BKS theory and makes it clear that the proposed mechanism is incompatible with strict energy conservation. According to Slater, the ``one \ldots essentially new" suggestion of BKS (note that he does not claim credit for the concept of virtual oscillators) was:
\begin{quotation}
that the wavelets sent out by an atom in connection with a given transition were sent out, not as a consequence of the occurrence of the transition, but as a consequence of the existence of the atom in the stationary state from which it could make that transition.\footnote{Note the similarity with the comments of Bohr to Ladenburg quoted in sec.\ 3.3:  ``the quantum jumps are not the direct cause of the absorption of radiation, but \ldots represent an effect  which accompanies the continuously disperging (and absorbing) effect of the atom on the radiation" \citep[Vol.\ 5, p.\ 400]{Bohr 1972--1996}.} On this assumption, the stationary state is the time during which the atom is radiating or absorbing; the transition from one state to another is not accompanied by radiation, but so far as the field is concerned, merely marks the end of the radiation or absorption characteristic of one state, and the beginning of that characteristic of another. The radiation emitted or absorbed during the stationary state is further not merely of the particular frequency connected with the transition which the atom is going to make; it includes all the frequencies connected with all the transitions which the atom could make \ldots Although the atom is radiating or absorbing during the stationary states, its own energy does not vary, but changes only discontinuously at transitions \ldots It is quite obvious that the mechanism becomes possible only by discarding conservation (ibid., pp.\ 397--398).
\end{quotation}
On the next page, Slater inserts a disclaimer similar to the one by Van Vleck  quoted above:
\begin{quotation}
It must be admitted that a theory of the kind suggested has unattractive features; there is an apparent duplication between the atoms on the one hand, and the mechanism of oscillators producing the field on the other. But this duplication seems to be indicated by the experimental facts, and it is difficult at the present stage to see how it is to be avoided (ibid., p.\ 399). 
\end{quotation}
Slater's portrayal of the BKS theory agrees with the exposition given by Kramers and Helge Holst (1871--1944) in the German edition \citep{Kramers and Holst 1925} of a popular book on Bohr's atomic theory originally published in Danish \citep{Kramers and Holst 1922}.\footnote{The translation was done by Fritz Arndt (1885--1969), a chemist and a colleague of Ladenburg and Reiche in Breslau (see the correspondence between Kramers and Ladenburg of 1923--1925 in the AHQP). The preface of this translation is dated March 1925.} In a section, entitled ``Bohr's new conception of the fundamental postulates," that was added to the German edition, Kramers explained that the BKS theory breaks with one of the basic tenets of Bohr's original theory, namely that atoms only emit light when one of its electrons makes a transition from, to use his example, the second to the first stationary state. ``According to the new conception," Kramers wrote, ``radiation with frequency $\nu_{\rm 2-1}$ is still tied to the {\it possibility} of a transition to the first state, but it is assumed that the emission  takes place during the entire time the atom is in the second state" \citep[p.\ 135]{Kramers and Holst 1925}. Another difference is that ``if the atom is in the third state, it will simultaneously emit the frequencies $\nu_{\rm 3-2}$ and $\nu_{\rm 3-1}$ until it either jumps to the second or to the first state" (ibid.). Kramers emphasizes  that this makes the new conception preferable to Bohr's original one from the point of view of the correspondence principle:
\begin{quotation}
This situation shows that the new conception is closer to the classical electron theory than the old one; the simultaneous emission of two frequencies mentioned above has its counterpart in that an electron moving on an ellipse emits both its fundamental tone and its first overtone \ldots while earlier one had to assume that these two frequencies were produced by {\it different} transitions in {\it different} atoms. Especially from the point of view of the correspondence principle is it a pleasure to welcome the fact that the radiation emitted by a {\it single} atom contains all the frequencies that correspond to possible transitions; for in the border region of large quantum numbers the radiation demanded by the quantum theory will now merge very smoothly with the radiation demanded by the classical theory \citep[pp.\ 135--136]{Kramers and Holst 1925}. 
\end{quotation}
The final paragraph of the BKS paper itself, from which we quoted above, can be seen as a garbled version of Kramers' argument here. Note that the term `virtual radiation' is absent from these expositions by Slater and Kramers.\footnote{In his detailed critique of the phyics of BKS, \citet{Dresden} struggles mightily to make sense of the ``somewhat vague, tenuous relation between the virtual field and the real electromagnetic field" (p.\ 179)}

\subsection{The demise of BKS}

The BKS theory was decisively refuted in experiments by Walther Bothe (1891--1957) and Hans Geiger (1882--1945) in Berlin and by Compton and Alfred Walter Simon in Chicago. These experiments showed that energy-momentum is strictly conserved in Compton scattering (i.e., event by event) and not just statistically (Stuewer, 1975, pp.\ 299--302; Stolzenburg, 1984, pp.\ 75--80). The experiments were begun shortly after the BKS paper was published (see Bothe and Geiger, 1924), but the final verdict did not come in until the following year. \citet{Bothe and Geiger 1925a, Bothe and Geiger 1925b} published their results in April 1925. The paper by \citet{Compton and Simon 1925} is signed June 23, 1925, and appeared in September 1925.\footnote{\citet[p.\ 301]{Stuewer 1975} draws attention to a footnote in this paper that makes it clear that the experiment had been discussed even before Slater's arrival in Copenhagen: ``The possibility of such a test was suggested by W.\ F.\ G.\ Swann in conversation with Bohr and one of us [Compton] in November 1923" \citep[p.\ 290, note 6]{Compton and Simon 1925}. Swann, the reader may recall, had just started in Chicago that fall, leaving the vacancy in Minnesota that was filled by Breit and Van Vleck (see sec.\ 2.2).} On April 17, 1925, Geiger sent Bohr a letter forewarning him of the results of his experiments with Bothe. When Geiger's letter arrived in Copenhagen four days later, Bohr was in the process of writing to Ralph H. Fowler (1889--1944) in Cambridge. In the postscript to this letter, Bohr conceded that  ``there is nothing else to do than to give our revolutionary efforts as honourable a funeral as possible" \citep[p.\ 301]{Stuewer 1975}. His co-authors Kramers and Slater took the fall of BKS harder. So did other supporters of the theory, such as Ladenburg, Reiche, and Born. By contrast, Einstein and Pauli, the theory's most vocal critics, rejoiced. As we shall see, Born, Pauli, and Van Vleck all explicitly recognized that the demise of BKS did not affect Kramers' dispersion theory and its virtual oscillators.  

Ladenburg and Reiche had first read (the German version of) the BKS paper \citep{BKS2} in May 1924. ``We are pleased," Ladenburg wrote to Kramers, ``that our considerations harmonize so well with your ideas."\footnote{Ladenburg to Kramers, May 31, 1924 (AHQP).} In the same letter, Ladenburg invited Kramers to come to Breslau to give a talk and to discuss in person what the two of them and Reiche had been discussing in correspondence (see sec.\ 3.4). Kramers accepted the invitation and suggested he talk about the new radiation theory, ``which, I hope, will soon meet with approval from most physicists (although I heard that Einstein has expressed a relatively unfavorable opinion)."\footnote{Kramers to Ladenburg, June 5, 1924 (AHQP).} Less than a week later, Kramers received the following intelligence from Ladenburg, directly addressing his parenthetical remark:
\begin{quotation}
As far as Einstein's opinion about your new conception of radiation is concerned, I can give you a very precise report, since I attended his talk on May 28 in the Berlin colloquium. His opinion was decidedly not unfavorable. He declared the new conception to be internally fully consistent and not in direct contradiction with any facts. The mechanism of the undulatory theory would have to be preserved in his opinion. He put great emphasis, however, on the conceptual logical difficulties of the new theory, of the ``preestablished harmony," which the fundamental introduction of probability instead of causality brings with it. Specific objections that he raised seemed to rest only on a not yet complete knowledge of all your considerations. He pointed to the asymmetry, for instance, that the production of virtual radiation was tied to a specific atomic state. In discussion, I pointed out in response to that that the virtual oscillators have the frequencies of {\it possible} transitions---at which point he immediately withdrew the objection.\footnote{Ladenburg to Kramers, June 8, 1924 (AHQP).}
\end{quotation} 
Privately, Einstein was less guarded. A month earlier---in a letter to Born and his wife Hedi (1892--1972) of April 29, 1924---he had already delivered his oft-quoted put-down that, should BKS turn out to be correct, he ``would rather have been a shoemaker or even an employee in a gambling casino than a physicist"  \citep[p.\ 32]{Klein 1970}.\footnote{For further discussion of Einstein's objections to BKS, see \citep[pp.\ 32--35]{Klein 1970}, \citep[pp.\ 255--263]{Wasserman}, and \citep[pp.\ 24--28, pp.\ 31--34]{Stolzenburg 1984}.} Talking to Kramers in late June, Einstein expressed himself more diplomatically again. Kramers stopped in Berlin on his return trip from Breslau, where he had given a well-received talk on BKS on June 24, 1924. As he reported to Ladenburg once he was back in Copenhagen: ``It was very interesting to hear Einstein's considerations; as he himself says, they are all arguments based on intuition."\footnote{Kramers to Ladenburg, July 3, 1924 (AHQP).}

Ladenburg also attended a colloquium in Berlin in May 1925 in which Bothe and Geiger presented their results. Ladenburg had just received a copy of the German edition of Kramers' popular book with Holst from which we quoted above. He clearly had a hard time accepting the refutation of BKS at this point. Referring to the discussion of BKS in ch.\ 6 of \citep{Kramers and Holst 1925}, he wrote:
\begin{quotation}
In this connection, I must report to you that yesterday Geiger and Bothe presented their important and beautiful experiments on counting electrons and [light] quanta in the Compton effect. Apparently, as you know, they have shown that the emission of electrons and quanta is simultaneous within one-thousandth of a second or less. Can I ask you to what extent you and Bohr consider this as standing in contradiction to your theory? Does your theory really require the complete independence of these two processes, so that only chance could cause the simultaneous occurrence of the two processes within one-thousandth of a second? You can imagine how these questions also affect us and if you have time to write to me to give your opinion I would be very grateful.\footnote{Ladenburg to Kramers, May 15, 1925 (AHQP).}
\end{quotation}
Unfortunately, we do not know whether and, if so, how Kramers replied.

When Slater found out about the experimental refutation of the BKS theory, he sent another letter to {\it Nature} (dated July 25, 1925) announcing that he had once more changed his mind:  ``The simplest solution to the radiation problem then seems to be to return to the view of a virtual field to guide corpuscular quanta" \citep{Slater 1925c}. Kramers and Bohr concurred: ``we think that Slater's original hypothesis contains a good deal of truth."\footnote{Kramers to Urey, July 16, 1925, quoted by \citet[p.\ 86]{Stolzenburg 1984}.} Slater thus reverted to the position that, as he reminds the reader, he had been talked out of by Bohr and Kramers. Slater also noted that Swann had argued for this view during the December 1924 meeting of the {\it American Association for the Advancement of Science}, unaware that he, Slater, had been thinking along the same lines.\footnote{Cf.\ \citep{Swann 1925}. See \citep[pp.\ 321--322]{Stuewer 1975} for discussion of Swann's proposal.} The following year, Bohr mentioned in passing in a letter to Slater that he had ``a bad conscience in persuading you to our view." Slater told him not to worry about it.\footnote{Bohr to Slater, January 28, 1926; Slater to Bohr, May 27, 1926 \citep[Vol.\ 5, pp.\ 68--69]{Bohr 1972--1996}.}

The way in which the BKS paper had come to be written, however, had left Slater with a bitter taste in his mouth \citep[pp.\ 350--356]{Schweber 1990}. 
We already quoted from his letter to Van Vleck of July 27, 1924, in which his disenchantment with Copenhagen shines through very brightly (see secs.\ 2.2 and 3.4). Interestingly, on that very same day, Slater wrote to Bohr, thanking him for his ``great kindness and attention to me while I was in Copenhagen. Even if we did have some disagreements, I felt very well repaid for my time there, and I look back to it very pleasantly" \citep[Vol.\ 5, p.\ 494]{Bohr 1972--1996}. This sounds disingenuous in view of his comments to Van Vleck, but Slater had also been very positive about Bohr writing to his teacher Bridgman on February 1, 1924 \citep[p.\ 354]{Schweber 1990}.
In his AHQP interview, however, Slater was very negative about Bohr and his institute. So negative, in fact, that when he found out that Copenhagen would be one of the depositories for the AHQP materials, he asked Kuhn to keep the interview out of the copy going to Denmark.\footnote{Slater to Kuhn, November 22, 1963, included in the folder on Slater in the AHQP.} 

Initially, Slater was angry with both Bohr and Kramers, but his attitude toward the latter later softened \citep[pp.\ 168--171]{Dresden}. This was probably under the influence of his wife, fellow-physicist Rose Mooney (1902--1981) \citep[pp.\ 527--528]{Dresden}.\footnote{A caveat is in order here. As John \citet{Stachel 1988} points out in his review of \citep{Dresden}, ``[t]he wealth of intimate detail about Kramers that Dresden provides relies so heavily on personal interviews (Dresden himself notes the ```soft' character" of this information) that it is difficult for others to assess the evidence until the interviews (which I hope were taped), as well as Kramers's personal papers, are made available to others" (p. 745).} Before Ms.\ Mooney became Mrs.\ Slater in 1948, she had been close to Kramers, whom she had met at a summer school in Michigan in 1938. The two of them almost certainly had an affair. Kramers was profoundly unhappy in his marriage to Anna `Storm' Petersen, a Danish singer he had met in artistic circles in Copenhagen and married in 1920 after she got pregnant.\footnote{Kramers was on the rebound at the time from the on-again-off-again relationship with his Dutch girlfriend, Waldi van Eck. Dresden's description of Kramers' relationship with van Eck (not to be confused with Van Vleck) conjures up the image of a virtual oscillator: ``no commitments were made, no decisions were taken, the relationship was never defined, it was certainly never consummated, nor ever terminated" \citep[p.\ 525]{Dresden}.} In one of the most memorable passages of his book, \citet[pp.\ 289--295]{Dresden} reveals that Kramers had told Storm many years after the fact that he himself had on at least one occasion been railroaded by Bohr. Kramers apparently thought of the Compton effect around 1920, well before Compton and Debye did. Bohr, however, detested the notion of light quanta so much that he worked on Kramers until he recanted. According to what Storm told Dresden, Kramers had to be hospitalized after one of these sessions with Bohr! Bohr's victory was complete. Even more strongly than Slater in the case of BKS a few years later, Kramers joined Bohr's crusade against light quanta with ``all the passion of a repentant convert" \citep[p.\ 171]{Dresden}. Slater may well have found out about this episode from his wife, Kramers' former mistress. Whether or not he did, in his autobiography, as \citet[p.\ 528]{Dresden} points out, \citet{Slater 1975} refers to his BKS co-author as ``my old friend Kramers" (p.\ 233).

Born had also been a supporter of BKS. With only Kramers' {\it Nature} notes to go on, he assumed that Kramers' dispersion theory was a product of BKS. He had no way of knowing that Kramers had these results before BKS. By the time \citep{Born 1924b} was published, however, Born realized that one did not have to subscribe to all articles of the BKS philosophy to extend the results of Kramers' dispersion theory. At the beginning of the paper, Born still writes as if the two theories stand or fall together:
\begin{quotation}
Recently \ldots considerable progress has been made by Bohr, Kramers and Slater on just this matter of the connection between radiation and atomic structure \ldots How fruitful these ideas are, is also shown by Kramers' success in setting up a dispersion formula \ldots In this situation, one might consider whether it would not be possible to extend Kramers' ideas, which he applied so successfully to the interaction between radiation field and radiating electron, to the case of the interaction between several electrons of an atom \ldots The present paper is an attempt to carry out this idea \citep[pp.\ 181--182]{Born 1924b}.
\end{quotation}
A footnote appended to this passage reads: ``By a happy coincidence I was able to discuss the contents of this paper with Mr.\ Niels Bohr, which contributed greatly to a clarification of the concepts." Bohr had visited Born and Heisenberg in G\"{o}ttingen in early June 1924 \citep[pp.\ 177--179]{Cassidy}. Heisenberg had already told Born all about BKS and Born had expressed his enthusiasm for the theory in a letter to Bohr of April 16, 1924.\footnote{See \citep[Vol.\ 5, p.\ 299]{Bohr 1972--1996}, discussed in \citep[Vol.\ 2, p.\ 143]{Mehra Rechenberg}.} Bohr's visit must have further solidified this enthusiasm. A week later, however, Einstein passed through G\"{o}ttingen and trashed the BKS theory.\footnote{See Heisenberg to Pauli, June 8, 1924 \citep[Doc.\ 62]{Pauli 1979}. This is the same day that Ladenburg wrote to Kramers that Einstein's opinion of BKS was ``decidedly not unfavorable" (see above).} As a result of Einstein's onslaught, Born hedged his bets and did not throw in his fate with the more controversial aspects of BKS (see Mehra and Rechenberg, 1982--2001, Vol.\ 2, p.\ 144; Cassidy, 1991, p.\ 179). At the beginning of sec.\ 3 of his paper, he writes:
\begin{quotation}
it will be profitable to make use of the intuitive ideas, introduced by Bohr, Kramers and Slater \ldots but our line of reasoning will be independent of the critically important and still disputed conceptual framework of that theory, such as the statistical interpretation of energy and momentum transfer \citep[p.\ 189]{Born 1924b}.\footnote{This illustrates the importance of what \citet{Beller} has called the ``dialogical approach" to the history of quantum mechanics (an approach adopted {\it avant la lettre} by Hendry [1984]): to resolve the tension between the two quoted passages in Born's paper, it is important to be attuned to the voices of both Bohr and Einstein in his text.}
\end{quotation}
Born, however, continued to be a true believer in BKS and took its collapse harder than Bohr himself. On April 24, 1925, he wrote to Bohr: 
\begin{quotation}
Today Franck showed me your letter [of April 21, 1925, the day that Bohr had received word from Geiger about the results of the Bothe-Geiger experiment] \ldots which interested me exceedingly and indeed almost shocked me, because in it you abandon the radiation theory that obeyed no conservation laws \citep[Vol.\ 5, p.\ 84]{Bohr 1972--1996}.
\end{quotation}
In contrast to Born, Pauli called the demise of BKS ``a magnificent stroke of luck."\footnote{Pauli to Kramers, July 27, 1925 \citep[pp.\ 232--234]{Pauli 1979} or \citep[Vol.\ 5, p.\ 87]{Bohr 1972--1996}.\label{pauli to kramers}} Pauli's opposition to BKS was probably fueled by Einstein, who gave him an earful about the theory during the annual meeting of the {\it Gesellschaft Deutscher Naturforscher und \"{A}rzte} in Innsbruck in September 1924.\footnote{See Pauli to Bohr, October 2, 1924 \citep[Doc.\ 66]{Pauli 1979}, quoted and discussed in \citep[pp.\ 260--263]{Wasserman}.} Pauli clearly recognized that Kramers' dispersion theory was independent of BKS and that the fall of the latter did not affect the former  \citep[p.\ 244]{Darrigol}. A footnote in \citep{Pauli 1925} emphasizes 
\begin{quotation}
that the formulas of [Kramers and Heisenberg, 1925] used here are independent of the special theoretical interpretation concerning the detailed description of the radiation phenomena in the quantum theory taken as a basis by them [i.e., the BKS theory], since these formulas only apply to averages over a large number of elementary phenomena \citep[p.\ 5]{Pauli 1925}.
\end{quotation}
As he explained to Kramers, Pauli wanted to distance himself from suggestion in the abstract of  \citep{Kramers and Heisenberg 1925} that ``the conclusions, should they be confirmed, would form an interesting support for this [i.e., the BKS] interpretation" (cf.\ sec.\ 4.1). Alerting Kramers to the footnote quoted above, Pauli wrote:
\begin{quotation}
if I had not added the footnote in question, it would also have been true that the conclusions of {\it my} paper, if they should be confirmed, `would form an interesting support for this interpretation.' This impression I had, of course, to counteract!\footnote{Pauli to Kramers, July 27, 1925 (cf.\ note \ref{pauli to kramers}).}
\end{quotation}
This letter was written after Heisenberg's {\it Umdeutung} paper, which was much more to Pauli's liking. In the same letter, in cruel Pauli fashion, he proceeded to berate Kramers for pushing the BKS theory. That this did not affect Pauli's appreciation for Kramers' work on dispersion is clear from what he wrote to another correspondent a few months after this scathing letter:  ``[m]any greetings also to Kramers, whom I am very fond of after all, especially when I think of his beautiful dispersion formula."\footnote{Pauli to Kronig, October 9, 1925, quoted in \citep[p.\ 91]{Stolzenburg 1984}}

In his NRC {\it Bulletin}, written after the Bothe-Geiger and Compton-Simon experiments, Van Vleck, like Pauli, stressed the independence of the Kramers dispersion theory and BKS. The rejection of the BKS theory and the acceptance of the light-quantum hypothesis, he wrote
\begin{quotation}
[do] not mean that Slater's concept of virtual oscillators is not a useful one. We may assume that the fields which guide the light-quants come from a hypothetical set of oscillators rather than from the actual electron orbits of the conventional electrodynamics.\footnote{At this point, the following footnote is appended: ``This viewpoint has been advocated by Slater during the printing of the present Bulletin. See [Slater, 1925a]."} In this way the appearance of the spectroscopic rather than the orbital frequency in dispersion can be explained, and the essential features of the virtual oscillator theory of dispersion \ldots can still be retained. There is an exact conservation of energy between the atoms and the actual corpuscular light-quants, but only a statistical conservation of energy between the atoms and the hypothetical virtual fields 
 \citep[pp.\ 286--287]{Van Vleck 1926a}. 
\end{quotation}
Virtual oscillators survived the demise of BKS and happily lived on in the dispersion theory from which they originated. 

These observations by Pauli and Van Vleck make it clear that BKS only played a limited role in the developments that led to matrix mechanics. It is important to keep that in mind. As long as we think of the Kramers dispersion theory as part and parcel of BKS, it looks as if matrix mechanics replaced a decisively refuted theory. Once we recognize that the Kramers dispersion theory was developed before and independently of BKS, we see that matrix mechanics grew naturally out of an eminently successful earlier theory. The BKS theory and its refutation by the Bothe-Geiger and Compton-Simon experiments then become a sideshow distracting from the main plot line, which runs directly from dispersion theory to matrix mechanics. A corollary to this last observation is that the acceptance of the light-quantum hypothesis was irrelevant to the development of matrix mechanics. Compton scattering provided convincing evidence for the light-quantum hypothesis and against the BKS theory, but it had no bearing on dispersion theory. This is not to deny that the light-quantum hypothesis indirectly played a role in dispersion theory: the work of Ladenburg, Kramers, and Van Vleck crucially depends on Einstein's $A$ and $B$ coefficients, Einstein's original derivation of which did involve light quanta.

\subsection{Heisenberg, BKS, and virtual oscillators}

When Heisenberg first read the BKS paper, he was not impressed: ``Bohr's paper on radiation is certainly very interesting; but I do not really see any fundamental progress."\footnote{Heisenberg to Pauli, March 4, 1924 \citep[Doc.\ 57]{Pauli 1979}; quoted by \citet[p.\ 202]{Dresden} and \citet[p.\ 250]{Wasserman}.} He subsequently warmed to the theory, writing to Copenhagen on April 6, 1924 that he hoped Bohr had meanwhile convinced Pauli.\footnote{See  \citep[Vol.\ 5, pp.\ 354--355]{Bohr 1972--1996}, cited by \citet[p.\ 176]{Cassidy} to support his claim that ``by the end of his March 1924 visit to Copenhagen, Werner was a convert."}  To Sommerfeld he wrote on November 18, 1924: ``Maybe Bohr's radiation theory is a most felicitous [{\it sehr gl\"{u}cklicke}] description  of this dualism [i.e., the wave-particle duality of radiation] after all" (Sommerfeld, 2004, p.\ 174, quoted in Wasserman, 1981, p.\ 251). Five years later, Heisenberg was praising BKS effusively:
\begin{quotation}
This investigation represented the real high point in the crisis of quantum theory, and, although it could not overcome the difficulties, it contributed, more than any other work of that time, to the clarification of the situation in quantum theory (Heisenberg, 1929, p.\ 492; translated and quoted in Stuewer, 1975, p.\ 291).
\end{quotation}
And thirty years later, \citet[p.\ 12]{Heisenberg 1955} remembered the BKS theory as ``the first serious attempt to resolve the paradoxes of radiation into rational physics" (quoted in Klein, 1970, p.\ 37).

Why was Heisenberg so taken with the BKS theory? We already came across part of the answer. As he told Kuhn in his AHQP interview, Heisenberg saw in BKS a precursor to the Born interpretation of the Schr\"odinger wave function (see sec.\ 4.1). This, we feel, mainly helps explain Heisenberg's profuse praise after the fact. In the same interview, however, Heisenberg identified another aspect of BKS that can account for his enthusiasm for BKS before {\it Umdeutung}---or rather, Kuhn identified it for him. What triggered Heisenberg's ruminations on probability in BKS and in the Born interpretation was the observation by Kuhn that despite the experimental refutation of the BKS theory, ``a large part of the basic ideas and the whole use of the Correspondence Principle formulated in terms of virtual oscillators goes on quite unshaken."\footnote{P.\ 2 of the transcript of session 4 of the AHQP interview with Heisenberg.} Heisenberg's response does not address this issue at all, whereupon Kuhn tries again: ``In order to do that paper [BKS] one talks not only about \ldots probability \ldots but also transforms one's idea of the atom into a collection of virtual oscillators that operate between states" (ibid., p.\ 3). This time Heisenberg takes the bait:
\begin{quotation}
Yes, that was it. This idea, of course, also was there already that an atom was really a collection of virtual oscillators. Now this \ldots was in some way contrary to the idea of an electron moving around a nucleus. The obvious connection, the only possible connection, was that the Fourier components of this motion in some way corresponded, as Bohr said, to the oscillators. But certainly this paper [BKS] then prepared the way for this later idea that the assembly of oscillators is nothing but a matrix. For instance, we can simply say that matrix elements are the collection of oscillators. In this way, you can say that matrix mechanics was already contained in this paper [BKS] (ibid., p.\ 3).
\end{quotation}
This supports the thesis in \citep{MacKinnon 1977} mentioned in sec.\ 4.2 that matrix mechanics can be seen as a theory of virtual oscillators. What we want to emphasize is that what initially seems to have attracted Heisenberg to the BKS theory was the notion of virtual oscillators. Given the origin of this concept, Heisenberg's intellectual debt on this point was not to BKS but---once again (see sec.\ 3.5)---to dispersion theory. During a subsequent session of the AHQP interview, Heisenberg, in fact, talks about the link between Fourier components and oscillators in the context Kramers' dispersion theory. ``When you say the dispersion formula started from a physical idea," Kuhn asked, ``do you have a particular thing in mind?" Heisenberg replied:
\begin{quotation}
Well, I would say that his [i.e., Kramers'] idea was that there was the Einstein paper [with the $A$ and $B$ coefficients] and there was the Ladenburg [1921] paper connected with Einstein's. On the other hand there was Bohr's Correspondence Principle and the idea finally that this has to do somehow with Fourier components as oscillators. Kramers had the force to combine these two possibilities in one simple formula---the dispersion formula. And this I think was a very important idea that one should combine the Einstein paper, which was very far from the Bohr model[,] with the Bohr model \ldots Behind this idea was already the idea of connecting the oscillators with the Fourier components, which, as I have said many times, was in the air somehow in these years.\footnote{P.\ 13 of the transcript of session 6 of the AHQP interview with Heisenberg.} 
\end{quotation} 
Heisenberg explicitly availed himself of virtual oscillators in \citep{Heisenberg 1925a}, a paper on the polarization of fluorescent light submitted from Copenhagen in November 1924 (i.e., before the Kramers-Heisenberg paper). Talking about this paper in his interview with Kuhn, Heisenberg said:
\begin{quotation}
I would say that all this is part of the game to make the total table of linear oscillators be the real picture of the atom. One felt that in the Correspondence Principle, one should compare one of these linear oscillators with one Fourier component of a motion \ldots So the whole thing was a program which one had consciously or unconsciously in one's mind. That is, how can we actually replace everywhere the orbits of the electron by the Fourier components and thereby get into better touch with what happens? Well, that was the main idea of quantum mechanics later on. One could see, more and more clearly, that the reality were the Fourier components and not the orbits.\footnote{P.\ 15 of the transcript of session 4 of the AHQP interview with Heisenberg. Parts of this passage are quoted in \citep[p.\ 155]{MacKinnon 1977} and in \citep[Vol.\ 2, p.\ 165]{Mehra Rechenberg} (although the latter cite their own conversations with Heisenberg as their source; cf.\ notes \ref{plagiarism1} and \ref{plagiarism2}).\label{plagiarism3}}
\end{quotation}
\citet[pp.\ 148--155]{MacKinnon 1977} stresses the importance of \citep{Heisenberg 1925a} for the development of matrix mechanics.\footnote{For other historical discussion of \citep{Heisenberg 1925a}, see \citep[pp.\ 187--188]{Cassidy} and \citep[Vol.\ 2, pp.\ 159--169]{Mehra Rechenberg}.} Heisenberg agreed. Commenting on a first draft of MacKinnon's article, he wrote to the author in July 1974: ``I was especially glad to see that you noticed how important the paper on the polarization of fluorescent light has been for my further work on quantum mechanics. Actually, in Copenhagen I felt that this paper contained the first step in which I could go beyond the views of Bohr and Kramers" \citep[p.\ 149, note 29]{MacKinnon 1977}. As he proudly recounts in his AHQP interview (see pp.\ 13--14 of the transcript of session 4), Heisenberg managed to convince Bohr and Kramers of his approach to this problem, an approach they initially questioned. 

\citet[pp.\ 157--162]{MacKinnon 1977} also sees \citep{Heisenberg 1925b} on the anomalous Zeeman effect as an important step on the way to matrix mechanics:
\begin{quotation}
In  the conclusion Heisenberg outlined a new program for quantum theory. One should use the virtual oscillator model to work out all the Fourier components for the electrons in an atom and for the coupling between electrons. In the rest of this article I will attempt to trace through in detail the way Heisenberg implemented this program and developed quantum mechanics \citep[pp.\ 161--162]{MacKinnon 1977}.
\end{quotation}
Here we part company with MacKinnon. Virtual oscillators are not mentioned at all in  \citep{Heisenberg 1925b} (though Fourier components are). \citet[p.\ 857]{Heisenberg 1925b} does not even refer to virtual oscillators when discussing results pertaining to incoherent radiation from \citep{Kramers and Heisenberg 1925}. This paper, far from being another step toward {\it Umdeutung}, seems to be mired in the intractable problems of the old quantum theory: the Zeeman effect, spin, and multi-electron atoms.

MacKinnon, in our opinion, thus overstates his case. Yet, even if we discard what he has to say about  \citep{Heisenberg 1925b} on the Zeeman effect, ample evidence remains for his claim that ``[t]he virtual oscillator model played an essential role in the process of reasoning that led Heisenberg to the development of quantum mechanics" \citep[p.\ 184]{MacKinnon 1977}. We also want to point out that the burden of proof for this thesis is not quite as heavy as MacKinnon makes it out to be. MacKinnon writes: ``after this paper [i.e., the {\it Umdeutung} paper] was written the virtual oscillator model sunk from sight and never resurfaced" (ibid). We already noted that the term ``substitute oscillators" can still be found  in the famous post-{\it Umdeutung} paper of \citet{Born and Jordan 1925b} (see sec.\ 3.3). What we did not mention so far is that  \citet[p.\ 456]{Lande 1926} actually introduced the phrase ``virtual orchestras" to describe not BKS but matrix mechanics!\footnote{Land\'{e} had worked with Heisenberg in 1924 \citep[p.\ 177]{Cassidy}, resulting in a joint paper \citep{Lande and Heisenberg}. In his AHQP interview,  Land\'{e} nonetheless said that \citep{Heisenberg 1925c} had been incomprehensible to him and that it had taken \citep{dreimaenner} for him to understand matrix mechanics (p.\ 3 of the transcript of session 5 of the interview; cf. note \ref{lande1}). These comments seem to be colored, however, by lingering resentment. Land\'{e} felt strongly that Born should have won the Nobel Prize for his contribution to matrix mechanics and that German antisemitism was the  only reason he had not.\label{lande2}} The imagery, if not exactly the language, of an ``orchestra of virtual oscillators" was also used in early popular expositions of matrix mechanics. In a popular book of the 1930s that went through many editions and was endorsed  by Max Planck in a short preface, Ernst Zimmer wrote:\footnote{We are grateful to J\"urgen Ehlers for drawing our attention to Zimmer's book.}
\begin{quotation}
The state of an atom should no longer be described by the unobservable position and momentum of its electrons, but by the measurable frequencies and intensities of its spectral lines \ldots Regardless of the nature of the real musicians  who play the optical music of the atoms for us, Heisenberg imagines assistant or auxiliary musicians [{\it Hilfsmusiker}]: every one plays just one note at a certain volume. Every one of these musicians is represented by a mathematical expression, $q_{mn}$, which contains the volume and the frequency of the spectral line as in expressions in acoustics familiar to physicists. These auxiliary musicians are lined up in an orchestra [{\it Kapelle}] according to the initial and final states $n$ and $m$ of the transition under consideration. The mathematician calls such an arrangement a ``matrix" \citep[pp.\ 161--162]{Zimmer 1934}. 
\end{quotation}
Zimmer's {\it Kapelle der Hilfsmusiker} was clearly inspired by Land\'e's  {\it Ersatz\-or\-chester der virtuellen Oszillatoren}. Virtual oscillators thus not only survived the demise of the BKS theory but also the transition to matrix mechanics. In fact, as we shall see in sec.\ 7.1, the features captured by the notion of virtual oscillators can still readily be identified in the formalism of modern quantum mechanics.
From the point of view of the quantum theory that emerged in the immediate aftermath of
Heisenberg's {\em Umdeutung} paper, in which the atomic system is quantized but not (as yet)  the ambient electromagnetic field, both the virtual oscillators and  the virtual radiation\footnote{Recall Heisenberg's comparison, quoted above, between the virtual field in BKS and the Schr\"odinger wave function.} of the BKS theory
are related not to the electromagnetic field but to the atomic system perturbed by this field. The effect of the perturbation is to induce additional Fourier components in the Schr\"odinger wave function of the electron. The virtual oscillators are just these harmonic components. Once the electromagnetic
field itself is quantized, it becomes more natural to identify the virtual oscillators
of BKS with the Fourier components of the quantized electromagnetic field, which correspond to 
time-dependent operators creating (or destroying) the photons emitted (or absorbed) by the atom.

\section*{Acknowledgments}
  
We are grateful to Soma Banerjee, Jeffrey Bub, Jeroen van Dongen, Michael Eckert, J\"{u}rgen Ehlers, Fred Fellows, Amy Fisher,  Clayton Gearhart, Domenico Giulini, Lee Gohlike, Seth Hulst, David Kaiser, Mary Kenney, Klaas Landsman, Christoph Lehner, John Norton, J\"urgen Renn, Serge Rudaz, Rob Rynasiewicz, Bob Seidel, Phil Stehle, Roger Stuewer, Kathreen Woyak, and Carol Zinda for comments, helpful discussion, and references. Earlier versions of parts of this paper were presented at Seven Pines VIII (Stillwater, MN, May 5--9, 2004), the Max Planck Institute for History of Science (Berlin, July 2004), New Directions in the Foundations of Physics (College Park, MD, April 29--May 1, 2005), the annual meeting of the History of Science Society (Minneapolis, MN, November 3--6, 2005), and at HQ0, a workshop on the history of quantum physics at the Max Planck Institute for History of Science (Berlin, June 13--16, 2006). The final version benefited greatly from comments by two anonymous referees. The authors gratefully acknowledge support from the Max Planck Institute for History of Science. The research of Anthony Duncan is supported in part by the National Science Foundation under grant PHY-0554660.

\end{document}